\documentclass[prl,twocolumn,superscriptaddress]{revtex4}
\usepackage{graphicx}
\usepackage{amsfonts}
\usepackage{amsmath}
\usepackage{amssymb}
\usepackage{bm}
\usepackage{slashed}
\usepackage{dsfont}

\usepackage{color}
\usepackage{soul}

\begin{document}
\title{Susceptibility indicator for chiral topological orders emergent from correlated fermions }
\author{Rui Wang}
\affiliation{National Laboratory of Solid State Microstructures and Department of Physics, Nanjing University, Nanjing 210093, China}
\affiliation{Collaborative Innovation Center for Advanced Microstructures, Nanjing University, Nanjing 210093, China}
\affiliation{Hefei National Laboratory, Hefei 230088, China}
\author{Tao Yang}
\affiliation{National Laboratory of Solid State Microstructures and Department of Physics, Nanjing University, Nanjing 210093, China}
\author{Z. Y. Xie}
\email{qingtaoxie@ruc.edu.cn}
\affiliation{Department of Physics, Renmin University of China, Beijing 100872, China}
\author{Baigeng Wang}
\email{bgwang@nju.edu.cn}
\affiliation{National Laboratory of Solid State Microstructures and Department of Physics, Nanjing University, Nanjing 210093, China}
\affiliation{Collaborative Innovation Center for Advanced Microstructures, Nanjing University, Nanjing 210093, China}
\author{X. C. Xie}
\affiliation{Hefei National Laboratory, Hefei 230088, China}
\affiliation{International Center for Quantum Materials, School of Physics, Peking University, Beijing 100871, China}
\affiliation{Collaborative Innovation Center of Quantum Matter, Beijing 100871, China}
\affiliation{CAS Center for Excellence in Topological Quantum Computation,
University of Chinese Academy of Sciences, Beijing 100190, China}

\begin{abstract}
Chiral topological orders formed in correlated fermion systems have been widely explored. However, the mechanism on how they emerge from interacting fermions is still unclear. Here, we propose a susceptibility condition. Under this condition, we show that chiral topological orders can spontaneously take place in correlated fermion systems. The condition leads to a low-energy effective theory of bosons with strong frustration, mimicking the flat band systems. The frustration then melts the long-range orders and results in topological orders with time-reversal symmetry breaking. We apply the theory to strongly-correlated semiconductors doped to the metallic phase. A novel excitonic topological order with semionic excitations and chiral excitonic edge state is revealed. We also discuss the application to frustrated magnets. The theory predicts a chiral spin liquid state, which is numerically confirmed by our tensor network calculations.  These results demonstrate an unprecedented indicator for chiral topological orders, which bridges the existing gap between interacting fermions and correlated topological matter.
\end{abstract}

\maketitle
\emph{\color{blue}{Introduction.--}} Chiral topological orders (TOs) breaking the time-reversal symmetry (TRS) have constituted a prominent topic over the last decades \cite{Kalmeyer,chiraspinstate}, and its discovery in fractional quantum Hall (FQH) systems \cite{xgwenc,laughlin,Tsui} invokes some of the most fundamental concepts in modern condensed matter physics \cite{xgwen1,witten1,Arovas,qniu,kitaev1}. Chiral topological orders are usually formed in correlated fermion systems. For example, the FQH states are generated from correlated electron states in Landau levels \cite{xgwenc,laughlin,Tsui}; the chiral spin liquids (CSLs) are formed in frustrated magnets \cite{yzhou,lbalents,YChe} or Mott insulators \cite{Bauer}, which again originate from strongly-correlated electronic materials. Moreover, the chiral excitonic topological order (ETO)  recently revealed in InAs/GaSb quantum wells also emerges from interacting electron-hole bilayers \cite{ruid}. These facts suggest there might be a underlying mechanism of chiral TOs accounting for their emergence from interacting fermions, which is yet to be addressed.

A convenient starting point to study interacting fermions is the Fermi liquid \cite{landauFL}. The instabilities of Fermi liquids provide a unified description of many long-range ordered states \cite{hvl,Honerkamp,cjwuu}, including superconductors, charge density waves, and magnetic orders, which can be understood as the condensation of bosons in corresponding channels. However, chiral TOs are disordered and characterized by long-range quantum entanglement, in stark contrast with the ordered states \cite{xgwensc,xchen,mlevin}. Therefore, it is of great challenge to find out their connections with the correlated fermion systems, which demands new developments beyond the conventional theory of Fermi liquid instability.

The two-dimensional (2D) interacting fermions is generally described by $H=H_0+H_I$, where $H_0=\sum_{\mathbf{r},\alpha}c^{\dagger}_{\mathbf{r}\alpha}h_{\alpha}(-i\nabla)c_{\mathbf{r},\alpha}$ and $h_{\alpha}(-i\nabla)=|\mathbf{k}|^2/2m_{\alpha}$ is the kinetic energy, with $\alpha=1,..,N$ being the band and spin indices. The fermion-fermion interaction reads as $ H_I=\sum_{\mathbf{r},\mathbf{r}^{\prime},\alpha,\beta}V(\mathbf{r}-\mathbf{r}^{\prime})c^{\dagger}_{\mathbf{r},\alpha}c_{\mathbf{r},\beta}c^{\dagger}_{\mathbf{r}^{\prime},\beta}c_{\mathbf{r}^{\prime},\alpha}$, where $\alpha\neq\beta$ is allowed. $\mathbf{g}=\{m_{\alpha},\mu,...\}$ denotes the model parameters, including the mass $m_{\alpha}$ and the chemical potential $\mu$. The key quantity indicating possible instabilities is the susceptibility $\chi_{\mathbf{g},\alpha\beta}(q)=-\sum_kG_{0,\alpha}(k)G_{0,\beta}(k+q)$ \cite{hvl,Honerkamp}, with $G_{0,\alpha}(k)$ the bare Green's function of fermions and $k=(\mathbf{k},i\omega_n)$. At the random phase approximation level, the interaction is renormalized as $V^{\prime}(q)=V(\mathbf{q})/[1+V(\mathbf{q})\chi_{\mathbf{g},\alpha\beta}(q)]$. It is known that, when the condition
$1+V(\mathbf{q})\chi_{\mathbf{g},\alpha\beta}(\mathbf{q},0)=0$ is satisfied at a single momentum point $\mathbf{q}=\mathbf{Q}$, the divergence indicates the formation of boson condensates or long-range orders.

In this work, we focus on an intriguing question, i.e., what is the fate of bosons if the above condition is simultaneously satisfied by an infinite number of points on a momentum loop, i.e.,
\begin{equation}\label{eq1}
    1+V(\mathbf{q})\chi_{\mathbf{g},\alpha\beta}(\mathbf{q},0)=0,~~~~\forall\mathbf{q}\in S^1,
\end{equation}
given $S^1$ a 1D loop embedded in 2D momentum space with the radius $Q$, as indicated by Fig.\ref{Fig1}(a).
Eq.\eqref{eq1} essentially implies that the bosons have an equal tendency to condense on each point of $S^1$, impying a strong frustration effect.  We show that Eq.\eqref{eq1} generates a low-energy effective theory describing interacting bosons on a moat-shaped band \cite{sur,Sedrakyan3,Sedrakyan1,Sedrakyan2}, as shown by Fig.\ref{Fig1}(b). The emergent physics mimics the flat band systems \cite{Regnault,xwan,zliu,dnsheng} along $S^1$ in a bosonic version, finally resulting in chiral TOs with TRS breaking. We apply the theory to study strongly-correlated semiconductors doped to the metallic phase, which satisfies the susceptibility condition in the particle-hole channel. A novel ETO state is revealed, which exhibits semionic anyons in the bulk and chiral excitonic edge state \cite{ruid}. More interestingly, by applying the theory to frustrated quantum magnets, we predict a chiral spin liquid, which is numerically confirmed by our tensor network calculations. These results reveal the long-desired connections between chiral TOs and interacting fermions.


\begin{figure}[t]
\includegraphics[width=1\linewidth]{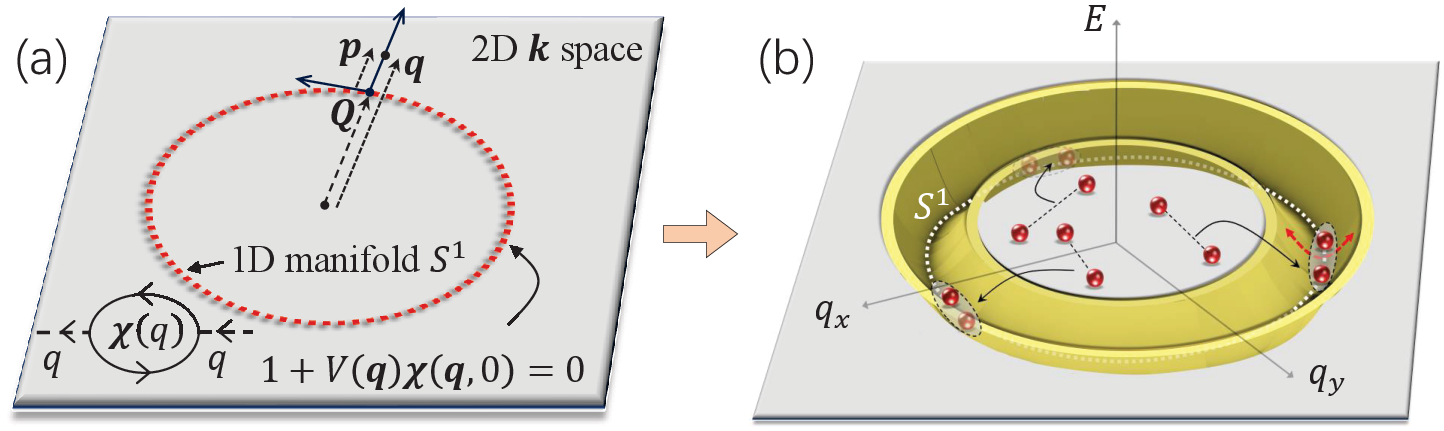}
\caption{(a) The correlated fermion systems satisfying the condition in Eq.\eqref{eq1} on a 1D manifold $\Lambda=S^1$ embedded in the 2D $\mathbf{k}$-space.  (b) Under  Eq.\eqref{eq1}, a low-energy effective theory takes place, which describes the fermion pairs on a moat-shaped band with the energy minima on $S^1$.  }\label{Fig1}
\end{figure}

\textit{\color{blue}{Emergence of Chiral TO under the susceptibility condition.}}-- We assume Eq.\eqref{eq1} is satisfied at $\mathbf{g}=\mathbf{g}_{cri}$ and study the general fermion model. By introducing the auxiliary bosons $O_{\mathbf{r},\alpha\beta}$, $H_I$  can be decomposed into the boson-fermion interaction:
\begin{equation}\label{eqhs}
  S_I=\int d\tau\sum_{\mathbf{r},\mathbf{r}^{\prime}}[O^{\dagger}_{\mathbf{r},\alpha\beta}V(\mathbf{r}-\mathbf{r}^{\prime})c^{\dagger}_{\mathbf{r}^{\prime},\beta}c_{\mathbf{r}^{\prime},\alpha}+h.c.]+...\\
\end{equation}
where the repeated indices are summed and ``..." denotes the boson bilinear terms. Introducing the operator $b_{\mathbf{r},a}\equiv b_{\mathbf{r},\alpha\beta}=\sum_{\mathbf{r}^{\prime}}V(\mathbf{r}-\mathbf{r}^{\prime})O_{\mathbf{r}^{\prime},\alpha\beta}$, where $a$ is the boson flavor with $a=1,...,N^2$, and integrating out the fermions \cite{sup,Melo}, we obtain $ Z=\int Db^{\star}Dbe^{-S_{eff}}$. The effective action of the bosons $S_{eff}$ reads as,
\begin{equation}\label{eqaction}
\begin{split}
  S_{eff}=-\mathrm{Trln}[-G^{-1}(\tau,\mathbf{r})]-\mathrm{Tr}[b^{\dagger}_{\mathbf{r}}V^{-1}(\mathbf{r}-\mathbf{r}^{\prime})b_{\mathbf{r}^{\prime}}],
\end{split}
\end{equation}
where $b_{\mathbf{r}}$ is the matrix with entries $b_{\mathbf{r},\alpha\beta}$.
$G^{-1}(\tau,\mathbf{r})=G^{-1}_0(\tau,\mathbf{r})-\Sigma(\tau,\mathbf{r})$ is the renormalized Green's function, with $ G^{-1}_0(\tau,\mathbf{r})=[-\partial_{\tau}-h(-i\nabla)]$.  $\Sigma(\tau,\mathbf{r})=b_{\mathbf{r}}+b^{\dagger}_{\mathbf{r}}$ is the self energy, which can be treated perturbatively at $\mathbf{g}=\mathbf{g}_{cri}$ \cite{Melo}.

To the second order perturbation, the action of bosons is obtained as $S_{0,eff}=\sum_a S^{(a)}_{0,eff}$, and the $a$-flavor sector is described by
\begin{equation}\label{eqSeffa}
  S^{(a)}_{0,eff}=-\sum_{q}[\chi_{\mathbf{g},a}(q)+V^{-1}(\mathbf{q})]b^{\dagger}_{q,a}b_{q,a}.
\end{equation}
The saddle point equation, $\delta S^{0,(a)}_{eff}/\delta b_{q,a}=0$, can be derived, which exactly reproduces the identity in Eq.\eqref{eq1}. Thus, Eq.\eqref{eq1} states that there are an infinite number of saddle point solutions and quantum fluctuations are non-negligible. Thus, we take into account the long-wave fluctuations around the saddle points. A generic momentum $\mathbf{q}$ can be measured in a local Cartesian coordinates with the origin $\mathbf{Q}$ and the unit vector tangential to $S^1$ (Fig.\ref{Fig1}(a)), i.e., $\mathbf{p}=\mathbf{q}-\mathbf{Q}$.
Then, making expansion with respect to $|\mathbf{p}|$ and $i\nu_n$ leads to the low-energy effective theory \cite{sup}:
\begin{equation}\label{eqgassian}
  S^{(a)}_{0,eff}=\sum_{i\nu_n,\mathbf{q}}[-i\nu_n+\frac{(|\mathbf{q}|-Q)^2}{2\tilde{m}_{a}}-\tilde{\mu}_{a}]b^{\dagger}_{q,a}b_{q,a},
\end{equation}
where $\tilde{m}_a$, $\tilde{\mu}_a$ are the effective mass and chemical potential of the $a$-flavor bosons, which are dependent on $\mathbf{g}$.
Interestingly, we observe from Eq.\eqref{eqgassian} that the kinetic energy of bosons is minimized for $|\mathbf{q}|=Q$, namely, on the loop $S^1$. Besides, quadratic dispersion takes place for $\mathbf{q}$ deviating from $S^1$, leading to the moat band shown in Fig.\ref{Fig1}(b). For $\mathbf{g}=\mathbf{g}_{cri}$ where Eq.\eqref{eq1} is satisfied, $\tilde{\mu}_a=0$ can be rigorously proved \cite{sup}.  This is a reflection of the fact that the bosons are about to condense at $\mathbf{g}=\mathbf{g}_{cri}$. However, as will be clear in the following, the condensation will be suppressed by quantum fluctuation.



To the fourth order, similar derivations as above  lead to the following effective action:
\begin{equation}\label{eqbosonint}
  S^{(a)}_{I,eff}=U_a\sum_{q_1,q_2,q_3}b^{\dagger}_{q_1,a}b_{q_1-q_2+q_3,a}b^{\dagger}_{q_3,a}b_{q_2,a},
\end{equation}
The coupling constant $U_a=4\sum_kG^2_{0,\alpha}(k)G^2_{0,\beta}(k)$ is generally positive.  Collecting both Eq.\eqref{eqgassian} and Eq.\eqref{eqbosonint}, we arrive at the effective Hamiltonian of bosons
\begin{equation}\label{eqGL}
  H_{eff}=\sum_{\mathbf{q}}[\frac{(|\mathbf{q}|-Q)^2}{2\tilde{m}}-\tilde{\mu}]b^{\dagger}_{\mathbf{q}}b_{\mathbf{q}}+U\sum_{\mathbf{r}}b^{\dagger}_{\mathbf{r}}b_{\mathbf{r}}b^{\dagger}_{\mathbf{r}}b_{\mathbf{r}},
\end{equation}
where the flavor $a$ is implicit. $\tilde{\mu}$ can be formally cancelled by shifting the zero of energy.  Clearly, Eq.\eqref{eqGL} describes interacting bosons on the moat band, as indicated by Fig.\ref{Fig1}(b).  

We now examine the possible ground state of Eq.\eqref{eqGL}. Due to the flatness of the moat band along $S^1$, the interaction $U$ plays the dominant role. In this case, the system can lower the energy cost from $U$ by statistical transmutations via the flux attachment \cite{kyang,kyangb,alopez}.  Technically, we represent the bosons as the composite fermions (CFs)  attached to an 1-flux quantum \cite{Jain,Halperin,SGK2,Ruia,Tigrana,ruib,ruic}, i.e., $\Psi_b(\mathbf{r}_1,...,\mathbf{r}_N)=\Psi_f(\mathbf{r}_1,...,\mathbf{r}_N)e^{i\sum_{i<j}\mathrm{arg}[\mathbf{r}_i-\mathbf{r}_j]}$. Although the fluxes cost the kinetic energy $\langle\Psi_b|H_K|\Psi_b\rangle$, where $H_K=(|\mathbf{q}|-Q)^2/2\tilde{m}$, the interaction energy from $U$, which plays the dominant role here, is significantly reduced in such a representation because of the antisymmetric nature of $\Psi_f(\mathbf{r}_1,...,\mathbf{r}_N)$. Hence, the fluxes that break TRS could be spontaneously generated, as they can further lower the system energy.


The next step is to look for the ground state wave function, $\Psi_f(\mathbf{r}_1,...,\mathbf{r}_N)$. Using $\Psi_b(\mathbf{r}_1,...,\mathbf{r}_N)=\Psi_f(\mathbf{r}_1,...,\mathbf{r}_N)e^{i\sum_{i<j}\mathrm{arg}[\mathbf{r}_i-\mathbf{r}_j]}$, the Hamiltonian in the fermionic basis describes composite fermions coupled to the Chern-Simons (CS) flux $B_{CS}=2\pi n$ \cite{ruic,Ruia,ruib,Tigrana}, where  $n$ is the fermion density \cite{footnote4}. Consequently, Landau quantization is formed, and the ground state in terms of $\Psi_f(\mathbf{r}_1,...,\mathbf{r}_N)$ is obtained to be the lowest Landau level state. Therefore, we arrive at the bosonic ground state wave function \cite{sup},
\begin{equation}\label{eqGS}
  \Psi_b({\mathbf{r}_1,...,\mathbf{r}_N})=\frac{1}{\sqrt{N!}}\mathrm{det}_{m,j}[\chi^l_m(z_j)]e^{i\sum_{i<j}\mathrm{arg}[\mathbf{r}_i-\mathbf{r}_j]},
\end{equation}
where $\chi^l_m(z_j)=A_{l,m}(\frac{z}{l_B})^me^{\frac{|z|^2}{4l^2_B}}L^{(m)}_l[\frac{|z|^2}{2l^2_B}]$ is the eigenstate of the $l$-th Landau level ($l$ determined by $n$ \cite{sup}) with the normalization factor $A_{l,m}$ and the complex coordinate $z$. Eq.\eqref{eqGS}  describes the lowest Landau level fully filled by composite fermions, which are bosons attached to 1-flux quanta.

The state in Eq.\eqref{eqGS} is essentially a chiral bosonic topological order, as it can be equivalently understood as an $\nu=1/2$ bosonic FQH \cite{sup} due to the following reason. Starting form an $\nu=1/2$ bosonic FQH under an intrinsic CS field $B_{\perp}$ and $\nu=\phi_D\rho_0/B_{\perp}=1/2$, where $\phi_D$ is the flux quantum and $\rho_0$ the particle number, and regarding the boson as the composite fermion attached to 1-flux quantum, then the effective field seen by the composite  fermions is $B_{CS}=B_{\perp}-\phi_D\rho_0=B_{\perp}/2$. Hence, the fermion filling factor is $\nu_{eff}=\phi_D\rho_0/B_{eff}=1$, leading to the fully-filled Landau level state in Eq.\eqref{eqGS}.

The energy of the state in Eq.\eqref{eqGS} can then be evaluated via $\langle\Psi_b|H_K|\Psi_b\rangle$, leading to $E_{TO}=\langle \Psi_b |[\frac{(|\mathbf{q}|-Q)^2}{2\tilde{m}}]|\Psi_b\rangle=\frac{\pi^2 n^2}{2\tilde{m}Q^2}\mathrm{log}^2\frac{4n}{Q^2}$ \cite{sup,Sedrakyan3}. Remarkably, in the low-density regime $n\rightarrow0$, $E_{TO}$ has lower energy than that of all the condensates proposed to date \cite{sup}, including the Fulde-Ferrell-Larkin-Ovchinnikov (FFLO) ($E_{FF/LO}\propto n$ \cite{Fulde,Larkin}) and the fragmented condensate ($E_{frag}\propto n^{4/3}$ \cite{Gopalakrishnan}). Notably, the energetics obtained here is confirmed by a recent Monte Carlo simulation \cite{cwei}.


Last, we recall that $\tilde{\mu}=0$ has been proved in Eq.\eqref{eqgassian} for $\mathbf{g}=\mathbf{g}_{cri}$ \cite{sup} and $n\propto2\sqrt{\tilde{\mu}}$ \cite{ruid}. Thus, $n\rightarrow0$ is always satisfied for  $\mathbf{g}\simeq\mathbf{g}_{cri}$.  Therefore, we arrive at the key conclusion of this work, i.e., the chiral TO described by Eq.\eqref{eqGS} will always emerge as the possible ground state, at least in a finite parameter region around $\mathbf{g}\simeq\mathbf{g}_{cri}$ where Eq.\eqref{eq1} is satisfied.

\begin{figure}
\includegraphics[width=1\linewidth]{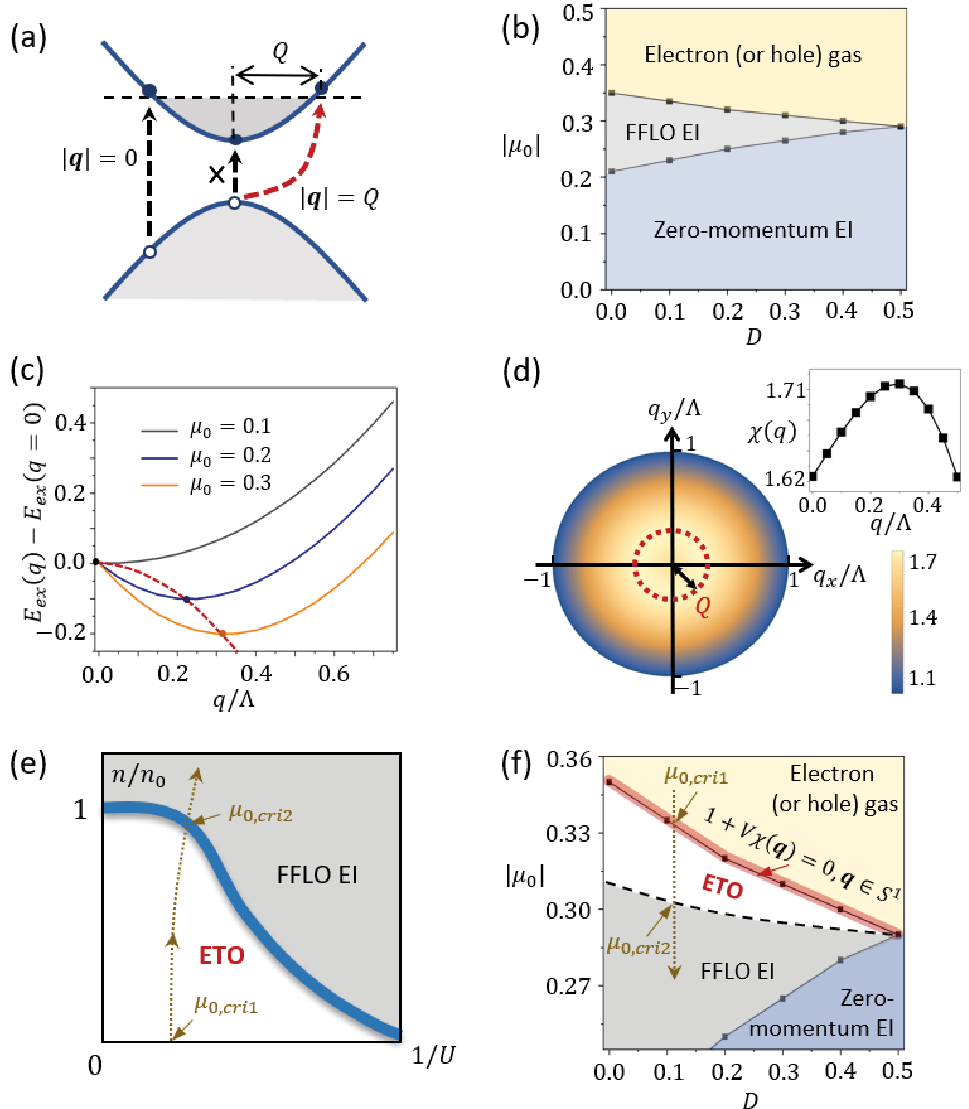}
\caption{(a) Plot of semiconductor doped to the metallic phase, with an implicit energy cutoff $W=\Lambda^2/2m$. The dashed arrows denote the excitation of electron-hole pairs. (b) The calculated mean-field phase diagram. $D$ and $|\mu_0|$ are in the unit of $W$. A strong interaction $V$ is required to obtain the EIs, and $V/W=1$ is used in (b). (c) The calculated exciton dispersion for different values of $\mu_0$ and $D=0.2$. The dashed red curve denotes the dispersion minimum with changing $\mu_0$.  (d) The density plot of the susceptibility in the particle-hole channel. The inset shows its dependence on $q$. (e) The phase diagram of the boson model in Eq.\eqref{eqGL}. (f) The zoom-in phase diagram obtained after considering the frustration effect. The parameters are the same with (b). The red thick curve denotes the critical regime  where Eq.\eqref{eq1} is satisfied. With varying the fermion parameters along the yellow dashed line, the parameters of boson theory in Eq.\eqref{eqGL} evolves along the yellow trajectory in (e). }\label{Fig2}
\end{figure}

\textit{\color{blue}{Appilication 1: excitonic topological order.}}-- We now apply the above general theory to study strongly-correlated doped semiconductors. The Hamiltonian is given by $H=H_0+H_I$, where $H_0=\sum_{\mathbf{k},n,\sigma}\epsilon_n(\mathbf{k})c^{\dagger}_{n,\mathbf{k},\sigma}c_{n,\mathbf{k},\sigma}$ with spin $\sigma$. The conduction and valence bands are described by $\epsilon_{\pm}(\mathbf{k})=\pm k^2/2m\pm D/2-\mu_0$, where $\mu_0$ is the chemical potential and $D$ is the band offset. We are interested in the doped metallic state with the electron (or hole) Fermi surface, i.e., $0<D<2|\mu_0|$, as indicated by Fig.\ref{Fig2}(a). On top of $H_0$, we consider the short-range inter-band interaction between the electrons, i.e., $ H_I=V\sum_{\mathbf{k},\mathbf{k}^{\prime},\mathbf{q}}\sum_{\sigma,\sigma^{\prime}}c^{\dagger}_{+,\mathbf{k}+\mathbf{q},\sigma}c_{+,\mathbf{k},\sigma}c^{\dagger}_{-,\mathbf{k}^{\prime}-\mathbf{q},\sigma^{\prime}}c_{-,\mathbf{k}^{\prime},\sigma^{\prime}}$. The intra-band interaction is negligible  as it only modifies the band dispersion via the mass renormalization.


Even if at low temperatures, there are virtual processes where electrons are excited to the conduction band above the Fermi level, leaving hole states in the valence band, as indicated by the dashed arrows in Fig.\ref{Fig2}(a). For strong $V$, the electrons and the holes can form excitons. When the binding energy overcomes the excitation energy, such virtual processes will be relevant, leading to condensation of excitons, i.e., excitonic insulators (EIs). In conventional mean-field theory \cite{sup}, we define the excitonic order parameter as $\Delta_{\tilde{\mathbf{q}}}=V\sum_{\mathbf{p}}\langle c^{\dagger}_{+,\mathbf{p},\sigma}c_{-,\mathbf{p}-\tilde{\mathbf{q}},\sigma}\rangle$ where $\tilde{\mathbf{q}}$ is the net exciton momentum. By  solving the mean-field equations, the order parameters $\Delta_{\tilde{\mathbf{q}}}$ is self-consistently determined. The solution $\Delta_{\tilde{\mathbf{q}}}=0$ simply describes the electron (or hole) gas without ordering. For the solutions where $\Delta_{\tilde{\mathbf{q}}}$ is maximized at $\tilde{\mathbf{q}}=0$, the excitons intend to exhibit zero net momentum. Such condensate is referred to the zero-momentum EI, in contrary to the condensate at finite momentum, i.e., the FFLO state.

The mean-field phase diagram with varying $|\mu_0|$ and $D$ is shown in Fig.\ref{Fig2}(b). As shown, in the strongly-doped regime with large $\mu_0$, the system remains as the electron (or hole) gas, because of the dominant excitation energy. For smaller $\mu_0$, the binding energy slightly overcomes the excitation energy. In this case, only the process indicated by the red dashed arrow in Fig.\ref{Fig2}(a) becomes relevant, as it has the lowest energy cost. Clearly, the excitons formed in this channel exhibit finite momenta, thus leading to the FFLO EI after condensation, as shown by Fig.\ref{Fig2}(b). With further lowering $|\mu_0|$, the binding energy becomes dominant, so that zero momentum excitons can be readily formed, as indicated by the $|\mathbf{q}|=0$ scattering channel in Fig.\ref{Fig2}(a). This leads to the zero-momentum EI in Fig.\ref{Fig2}(b).

We then examine the excitation energy as a function of the exciton momentum $\mathbf{q}$.  From Fig.\ref{Fig2}(a), we observe that it decreases from $|\mathbf{q}|=0$, and reaches the minimum at $|\mathbf{q}|=Q$. This fact is further manifested by the exciton energy $E_{ex}$, which can evaluated by minimizing  $W_{ex}=\langle FS|\Delta_{\mathbf{p}}[H-E_{ex}(\mathbf{p})] \Delta^{\dagger}_{\mathbf{p}}|FS\rangle$, where $\Delta^{\dagger}_{\mathbf{p}}=\sum_{\mathbf{k}}\phi_{\mathbf{p}}(\mathbf{k})c^{\dagger}_{+,\mathbf{k},\sigma}c_{-,\mathbf{p}-\mathbf{k},\sigma}$  with the variation parameter $\phi_{\mathbf{p}}(\mathbf{k})$, and $|FS\rangle$ denotes the Fermi sea \cite{sup}. As shown by Fig.\ref{Fig2}(c), for the metallic states, the calculated exciton dispersion exhibits the minimum at finite momentum, leading to the moat-like band of excitons, as that in Fig.\ref{Fig1}(b). This indicates that there emerges a frustrated ordering tendency along $S^1$.

We further calculate the susceptibility $\chi(\mathbf{q},0)$ in the particle-hole channel. As shown by Fig.\ref{Fig2}(e), $\chi(\mathbf{q},0)$  exhibits the same maxima along a momentum loop, as indicated by the red dashed circle. Hence, for a proper parameter $\mathbf{g}=\mathbf{g}_{cri}$,  $1+V(\mathbf{q})\chi(\mathbf{q},0)=0$ can be satisfied by all $\mathbf{q}\in S^1$, leading to correlated excitons on the moat band \cite{sup}. Hence, a chiral TO formed by excitons, i.e., the ETO, should emerge as ground state for $\mathbf{g}\sim\mathbf{g}_{cri}$ according to our theory.

Comparison of the energetics between the boson condensates and the ETO  leads to the phase diagram corresponding to Eq.\eqref{eqGL}, as shown in Fig.\ref{Fig2}(e). In terms of the original fermion model, we gradually decrease $\mu_0$ along the dashed line in Fig.\ref{Fig2}(f). For $\mu_0=\mu_{0,cri1}$, where the susceptibility condition is satisfied, $\tilde{\mu}\sim0$ in Eq.\eqref{eqGL}, thus the density $n\sim0$, as discussed above. With further decreasing $\mu_0$, $n$ increases and moves along the dashed trajectory in Fig.\ref{Fig2}(e). The system stays in the ETO state until the trajectory crosses the ETO-FFLO boundary at $\mu_0=\mu_{0,cri2}$. This determines the ETO region of the phase diagram in Fig.\ref{Fig2}(f) Thus, the quantum fluctuation results in a translational invariant, long-range quantum entangled chiral excitonic state\cite{ruid,ldu1,ldu2}.

\begin{figure}[t]
\includegraphics[width=1\linewidth]{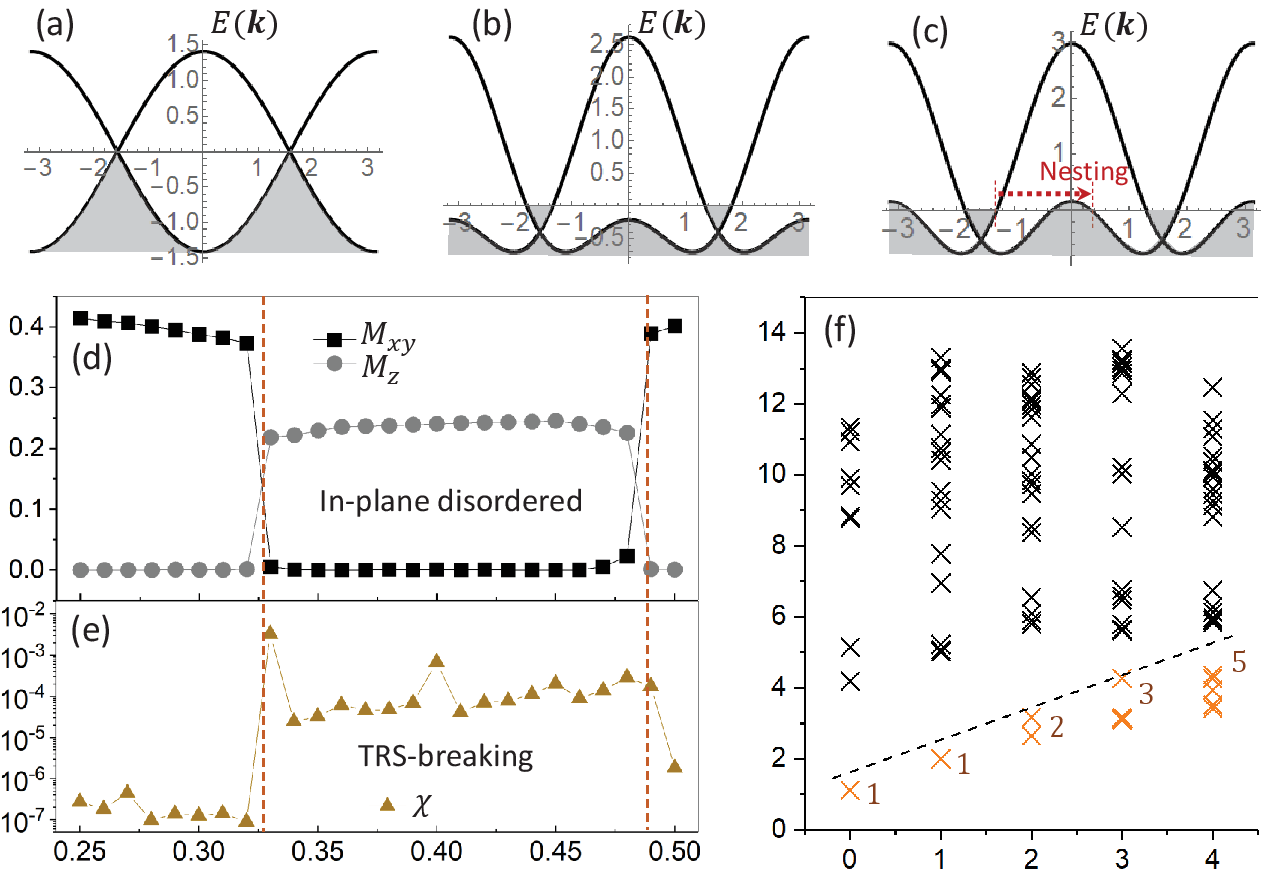}
\caption{(a-c) The dispersion of fermions along $k_x=k_y$ after 2D Jordan-Wigern transformation of the $J_1$-$J_2$-$J_3$ XY square model.  $J_2/J_1=0,0.4,0.52$ for (a), (b), and (c), respectively. (d) The calculated in-plane $M_{xy}$ and out-of-plane $M_z$ magnetization with increasing $J_2/J_1$. (e) The calculated chirality order $\chi=\langle\mathbf{S}_1\cdot(\mathbf{S}_2\times\mathbf{S}_3)\rangle$ with $\mathbf{S}_{1,2,3}$ the spin operator defined on three sites of a square plaquette. (f) The entanglement spectrum as a function of $k_y$. $J_2/J_1=0.4$ and $k_y$ is in unit of $2\pi/5$. The bond dimension is $D=10$ in the calculations. }\label{Fig3}
\end{figure}
\textit{\color{blue}{Application 2: chiral spin liquid and numerical evidences.}}-- Our theory can also be applied to predict chiral spin liquids in frustrated magnets. We consider  the spin-1/2 $J_1$-$J_2$-$J_3$ XY model on square lattice, which satisfies the susceptibility condition as will be shown below. The model is defined as $H=\sum_{\mathbf{r},\mathbf{r}^{\prime}}(J_{\mathbf{r},\mathbf{r}^{\prime}}/2)(S^x_{\mathbf{r}}S^x_{\mathbf{r}^{\prime}}+S^y_{\mathbf{r}}S^y_{\mathbf{r}^{\prime}})$ where the sum evolves the first, second, and third nearest neighbor bonds. We fermionize the quantum spin model using the 2D Jordan-Wigner transformation \cite{kyang,kyangb,alopez,Ruia,ruib,Tigrana}, which transform the spin operator into fermions and the lattice gauge field. The lattice gauge field has two effects. One is to generate fluxes that minimize the energy of fermions, and the other is to mediate fermion-fermion interactions.  Focusing on $J_3=J_2/2$ and gradually increasing $J_2$, we plot the single-particle dispersion of the fermions in Fig.\ref{Fig3}(a)-(c). As shown, the Dirac fermion states emerge, and the chemical potential $\mu_0$ gradually moves away from the Dirac points with increasing $J_2$.
 For small $J_2$, the gauge-mediated fermion-fermion interaction has been shown to induce paring of fermions, whose condensation leads to N\'{e}el order \cite{Tigrana,ruib}. While for large $J_2$, another Fermi pocket emerges around $\Gamma$ point (Fig.\ref{Fig3}(c)). Then, the Fermi surface nesting would favor spin density waves \cite{SGK2}. Interestingly, for intermediate $J_2$ (Fig.\ref{Fig3}(b)), the same low-energy physics as that of the ETO example (Fig.\ref{Fig2}(a)) occurs, and the susceptibility condition is satisfied \cite{sup}. Therefore, a chiral TO is expected to take place in between two magnetically ordered states.

We then use tensor network calculations \cite{PESS2014,SU1D2007,SU2D2008,FU2014,CTMRG1996,CTMRG2009,CTMRG2014,Cirac2011,Poil2016,Saeed2018,krylov,Verstraete,zyxieb} to simulate the ground state. As shown by Fig.\ref{Fig3}(d), in between two in-plane magnetic orders, an intermediate phase ($0.33\lesssim J_2/J_1\lesssim0.49$) occurs, which is completely free from any in-plane ordering. The chirality order also shows a significant enhancement in this region, clearly indicating the spontaneous breaking of TRS. Moreover, the entanglement spectrum exhibits the level counting, 1,1,2,3,5... (Fig.\ref{Fig3}(f)), in consistent with the $\mathrm{SU}(2)_1$ conformal field theory, implying the existence of the chiral edge state. These data offer strong evidences for the chiral spin liquid state, justifying our analytic predictions.



\textit{\color{blue}{Summary and discussion.}}--  The ETO is a chiral bosonic TO exhibiting semionic excitations in the bulk and chiral excitonic edge stated \cite{sup,ruid}. The experimental evidence was recently reported in the semimetal phase of InAs/GaSb quantum wells \cite{ruid,ldu1,ldu2} with density imbalance \cite{kunyang3}. Here, we reveal that ETO can even be formed in the metallic phase of doped semiconductors. In this case, a strong interaction $V$ comparable to the band width $W$ is desired. The twisted TMD bilayers provide a promising platform, which can realize semiconductors with strong correlation and remarkably flat bands \cite{fuliang}. Therefore, our theory could have intimate connections with the recently reported  fractional quantum anomalous Hall states in the twisted Mori\'{e} systems \cite{jcai,park,fxu,yacoby}.

The mechanism revealed here applies to correlated fermionic systems, in which the number of bosons depends on how many fermions are paired. In this case, the system can always lower its energy at the optimal density \cite{ruid} where the lowest Landau level is fully filled. In contrast, for bosons with fixed density on a moat band, generic filling of the Landau level is likely. Consequently, metallic states with quasi-long-range order are expected \cite{sur}.  The proposed mechanism may also be used to predict other chiral TOs, such as fractional Chern insulators \cite{zliu,Regnault}. The generalization to TRS-preserving TOs and non-Abelian TOs is also an interesting direction.


\begin{acknowledgments}
R. W. acknowledges Tigran Sedrakyan, Qianghua Wang, Rui-Rui Du, Lingjie Du and Y. X. Zhao  for fruitful discussions. This work was supported by the National R\&D Program of China (Grants No. 2023YFA1406500, 2022YFA1403601), the National Natural Science Foundation of China (No. 12322402, No. 12274206), the Innovation Program for Quantum Science and Technology (Grant no. 2021ZD0302800), and the Xiaomi foundation.
\end{acknowledgments}


\pagebreak
\vspace{5cm}
\widetext
\setcounter{equation}{0}
\setcounter{figure}{0}
\setcounter{table}{0}
\setcounter{page}{1}
\makeatletter
\renewcommand{\theequation}{S\arabic{equation}}
\renewcommand{\thefigure}{S\arabic{figure}}
\renewcommand{\bibnumfmt}[1]{[S#1]}
\renewcommand{\citenumfont}[1]{S#1}

\pagebreak
\vspace{5cm}
\widetext
\begin{center}
\textbf{\large Supplemental material for: Susceptibility indicator for chiral topological orders emergent from correlated fermions}
\end{center}




\date{\today }
\maketitle

\section{1. Emergent kinetic frustration for bosons in correlated fermions}
\subsection{1.1 Fermion-boson decomposition}
We consider the general interacting model given by $H=H_0+H_I$, where $H_0$ describes the free Hamiltonian, i.e.,
\begin{equation}\label{eq9}
  H_0=\sum_{\mathbf{r},\alpha\beta}c^{\dagger}_{\mathbf{r}\alpha}h_{\alpha,\beta}(-i\nabla)c_{\mathbf{r},\beta},
\end{equation}
where $\alpha,\beta=1,2,...N$ denotes the band index, spin, etc. $h_{\alpha\beta}(-i\nabla)$ is the single-particle Hamiltonian. We will focus on the diagonal case with $h_{\alpha\beta}=h_{\alpha}\delta_{\alpha\beta}$, and generalization to the off-diagonal cases are straightforward. We also consider the fermion-fermion interaction,
\begin{equation}\label{eq10}
  H_I=\sum_{\mathbf{r},\mathbf{r}^{\prime},\alpha,\beta}V(\mathbf{r}-\mathbf{r}^{\prime})c^{\dagger}_{\mathbf{r},\alpha}c_{\mathbf{r},\beta}c^{\dagger}_{\mathbf{r}^{\prime},\beta}c_{\mathbf{r}^{\prime},\alpha},
\end{equation}
where the flavor-exchange terms ($\alpha\neq\beta$) are allowed. The interaction can be decoupled via Hubbard-Stratonovich transformation to
\begin{equation}\label{eq11}
  -S_I=\int d\tau\sum_{\mathbf{r},\mathbf{r}^{\prime}}[O^{\dagger}_{\mathbf{r},\alpha\beta}V(\mathbf{r}-\mathbf{r}^{\prime})O_{\mathbf{r}^{\prime},\beta\alpha}-O^{\dagger}_{\mathbf{r},\alpha\beta}V(\mathbf{r}-\mathbf{r}^{\prime})c^{\dagger}_{\mathbf{r}^{\prime},\beta}c_{\mathbf{r}^{\prime},\alpha}-V(\mathbf{r}-\mathbf{r}^{\prime})c^{\dagger}_{\mathbf{r},\alpha}c_{\mathbf{r},\beta}O_{\mathbf{r}^{\prime},\beta\alpha}],
\end{equation}
where the sum of the repeated indices is implied. Note that the first term can also be written into the convolution, i.e.,
\begin{equation}\label{eq12}
  \sum_{\mathbf{r},\mathbf{r}^{\prime}}O^{\dagger}_{\mathbf{r},\alpha\beta}V(\mathbf{r}-\mathbf{r}^{\prime})O_{\mathbf{r}^{\prime},\beta\alpha}=\sum_{\mathbf{q}}O^{\dagger}_{\mathbf{q},\alpha\beta}V_{\mathbf{q}}O_{\mathbf{q},\beta\alpha}=\sum_{\mathbf{r},\mathbf{r}^{\prime},\mathbf{x},\mathbf{y}}O^{\dagger}_{\mathbf{r},\alpha\beta}V(\mathbf{r}-\mathbf{x})V^{-1}(\mathbf{x}-\mathbf{y})V(\mathbf{y}-\mathbf{r}^{\prime})O_{\mathbf{r}^{\prime},\beta\alpha}.
\end{equation}
Then we define the bosonic operators as
\begin{eqnarray}
  b_{\mathbf{r},\beta\alpha} &=& \sum_{\mathbf{r}^{\prime}} V(\mathbf{r}-\mathbf{r}^{\prime})O_{\mathbf{r}^{\prime},\beta\alpha}, \\
  b^{\dagger}_{\mathbf{r}^{\prime},\alpha\beta} &=& \sum_{\mathbf{r}}V(\mathbf{r}-\mathbf{r}^{\prime})O^{\dagger}_{\mathbf{r},\alpha\beta}.
\end{eqnarray}
Inserting the above two equations into Eq.\eqref{eq11} and Eq.\eqref{eq12}, we obtain
\begin{equation}\label{eq13}
  -S_I=\int d\tau\sum_{\mathbf{r},\mathbf{r}^{\prime}}b^{\dagger}_{\mathbf{r},\alpha\beta}V^{-1}(\mathbf{r}-\mathbf{r}^{\prime})b_{\mathbf{r}^{\prime},\beta\alpha}-\sum_{\mathbf{r}}b^{\dagger}_{\mathbf{r},\alpha\beta}c^{\dagger}_{\mathbf{r},\beta}c_{\mathbf{r},\alpha}-\sum_{\mathbf{r}}c^{\dagger}_{\mathbf{r},\alpha}c_{\mathbf{r},\beta}b_{\mathbf{r},\beta\alpha}].
\end{equation}
In addition to $S_I$, the action for the free part is given by
\begin{equation}\label{eq14}
  S_0=\int d\tau\sum_{\mathbf{r},\alpha\beta}c^{\dagger}_{\mathbf{r},\alpha}[\partial_{\tau}+h_{\alpha\beta}(-i\nabla)]c_{\mathbf{r},\beta}.
\end{equation}
It is convenient to introduce the fermionic spinor, $c_{\mathbf{r}}=[c_{\mathbf{r},1},c_{\mathbf{r},2},...,c_{\mathbf{r},3}]^\mathrm{T}$ and the matrix $b_{\mathbf{r}}$ in the flavor space. Then, the total action is cast into:
 \begin{equation}\label{eq15}
   S=\int d\tau\sum_{\mathbf{r}}c^{\dagger}_{\mathbf{r}}[\partial_{\tau}+h(-i\nabla)+(b^{\dagger}_{\mathbf{r}}+b_{\mathbf{r}})]c_{\mathbf{r}}-\int d\tau\sum_{\mathbf{r},\mathbf{r}^{\prime}}\mathrm{tr}[b^{\dagger}_{\mathbf{r}}V^{-1}(\mathbf{r}-\mathbf{r}^{\prime})b_{\mathbf{r}^{\prime}}],
 \end{equation}
where ``$\mathrm{tr}$" denotes the trace over the matrix. To further simplify the notations, we introduce the bare Matsubara Green's function as
\begin{equation}\label{eq16}
  G_0(\tau,\mathbf{r})=[-\partial_{\tau}-h(-i\nabla)]^{-1}
\end{equation}
and the self-energy matrix defined as
\begin{equation}\label{eq17}
  \Sigma(\tau,\mathbf{r})=b_{\mathbf{r}}+b^{\dagger}_{\mathbf{r}}.
\end{equation}
Then, the partition function $Z$ is cast into the functional integral over both the auxiliary field $b$ and the Grassmann field $c$, namely,
\begin{equation}\label{eq18}
  Z=\int D\overline{c}DcDb^{\star}Dbe^{-S},
\end{equation}
where $S$ is obtained as
\begin{equation}\label{eq19}
  -S=\int d\tau\sum_{\mathbf{r}}c^{\dagger}_{\mathbf{r}}[G^{-1}_0(\tau,\mathbf{r})-\Sigma(\tau,\mathbf{r})]c_{\mathbf{r}}+\int d\tau\sum_{\mathbf{r},\mathbf{r}^{\prime}}\mathrm{tr}[b^{\dagger}_{\mathbf{r}}V^{-1}(\mathbf{r}-\mathbf{r}^{\prime})b_{\mathbf{r}^{\prime}}].
\end{equation}
We formally integrate out the fermionic fields in Eq.\eqref{eq19}, which leads to
\begin{equation}\label{eq20}
  -S_b=\mathrm{Trln}[-G^{-1}(\tau,\mathbf{r})]+\int d\tau\sum_{\mathbf{r},\mathbf{r}^{\prime}}\mathrm{tr}[b^{\dagger}_{\mathbf{r}}V^{-1}(\mathbf{r}-\mathbf{r}^{\prime})b_{\mathbf{r}^{\prime}}],
\end{equation}
where $G^{-1}(\tau,\mathbf{r})=G^{-1}_0(\tau,\mathbf{r})-\Sigma(\tau,\mathbf{r})$ is the renormalized Green's function.

Our aim is to examine whether the boson condensation takes place. With tuning the parameter $\mathbf{g}$ implicit in the model, one assumes that the boson condensation is about to take place at a critical parameter $\mathbf{g}_{cri}$, then infinitesimal $\langle b_{\mathbf{r}}\rangle\rightarrow0^+$ would be developed near $\mathbf{g}_{cri}$. Thus, one can make expansion with respect to $\Sigma(\mathbf{r})$ in the logarithmic function in Eq.\eqref{eq20}. Up to the fourth-order expansion, this leads to a bosonic theory given by
\begin{equation}\label{eq21}
  S_{eff}=\frac{1}{2}\mathrm{Tr}(G_0\Sigma)^2+\frac{1}{4}\mathrm{Tr}(G_0\Sigma)^4
  -\int d\tau\sum_{\mathbf{r},\mathbf{r}^{\prime}}\mathrm{tr}[b^{\dagger}_{\mathbf{r}}V^{-1}(\mathbf{r}-\mathbf{r}^{\prime})b_{\mathbf{r}^{\prime}}],
\end{equation}
where ``$\mathrm{Tr}$" represents for the trace over the matrix in the flavor space, as well as the integral over spatial coordinates and the imaginary time.

We now calculate the first term in Eq.\eqref{eq21}, i.e.,
\begin{equation}\label{eq22}
  \frac{1}{2}\mathrm{Tr}(G_0\Sigma)^2=\frac{1}{2}\sum_{r,r^{\prime}} G_{0,\alpha\beta}(r-r^{\prime})\Sigma_{\beta\delta}(r^{\prime})G_{\delta\gamma}(r^{\prime}-r)\Sigma_{\gamma\alpha}(r),
\end{equation}
where we introduced $r=(\tau,\mathbf{r})$ for brevity. Making transformation to the momentum space and using $b^{\dagger}_{\mathbf{q}}=b_{-\mathbf{q}}$ (since $b_{\mathbf{q}}\propto\sum_{\mathbf{k}}c^{\dagger}_{\mathbf{k}}c_{\mathbf{k}+\mathbf{q}}$), Eq.\eqref{eq22} is reduced to:
\begin{equation}\label{eq23}
  \frac{1}{2}\mathrm{Tr}(G_0\Sigma)^2=\sum_{k,q}G_{0,\alpha\beta}(k)G_{0,\delta\gamma}(k+q)b^{\dagger}_{q,\gamma\alpha}b_{q,\beta\delta}.
\end{equation}
Thus, to the second-order expansion, we obtain the Gaussian level action,
\begin{equation}\label{eq24}
  S_{0,eff}=\sum_{k,q}G_{0,\alpha\beta}(k)G_{0,\delta\gamma}(k+q)b^{\dagger}_{q,\gamma\alpha}b_{q,\beta\delta}- \sum_{q}V^{-1}(\mathbf{q})b^{\dagger}_{q,\alpha\beta}b_{q,\beta\alpha}.
\end{equation}

Before proceeding, let us derive the saddle point equation by minimizing the action $S_{0,eff}$, i.e., $\delta S_{0,eff}/\delta b^{\star}_{q,\rho\sigma}=0$. This leads to
\begin{equation}\label{eq25}
  \delta S_{0,eff}/\delta b^{\star}_{q,\rho\sigma}=V^{-1}(\mathbf{q})b_{q,\sigma\rho}+\chi_{\beta\delta,\sigma\rho}(q)b_{q,\sigma\rho}=0,
\end{equation}
where we define the susceptibility as
\begin{equation}\label{eq26}
  \chi_{\beta\delta,\sigma\rho}(q)=-\sum_kG_{0,\beta\sigma}(k)G_{0,\rho\delta}(k+q).
\end{equation}
Treating the susceptibility as a matrix with the indices, $a\equiv(\beta,\delta)$ and $b\equiv(\sigma,\rho)$ (as defined in the main text), Eq.\eqref{eq25} can then be written into the compact matrix form as
\begin{equation}\label{eq27}
  [V^{-1}(\mathbf{q})\mathbb{I}+\boldsymbol{\chi}(q)]b_{\mathbf{q}}=0,
\end{equation}
where $\mathbb{I}$ is the identity matrix in the flavor space, i.e., $\mathbb{I}=\delta_{a,b}$, $\boldsymbol{\chi}(q)$ is the susceptibility matrix. For the case where only the bosons $b_{\sigma_0,\rho_0}$ are formed, Eq.\eqref{eq25} is reduced to
\begin{equation}\label{eq27add}
V^{-1}(\mathbf{q})+\chi_{\beta\delta}(\mathbf{q},0)=0
\end{equation}
where we have set $i\nu_n=0$ since the boson condensates are studied.  Clearly, the saddle point equation  Eq.\eqref{eq27add} produces the mean-field equation for the bosonic order parameter at the critical point $\mathbf{g}_{cri}$, and it describes the tendency of developing long-range orders.

Our main focus here is the 2D correlated systems with frustrated ordering tendency along a 1D manifold, thus we are interested in the following condition:
\begin{equation}\label{eq29}
  1+V(\mathbf{q})\chi_{\beta\delta}(\mathbf{q},0)=0,~~~~ \forall\mathbf{q}\in\Lambda,
\end{equation}
given $\Lambda$ a generic 1D manifold embedded in 2D momentum space. Note that the susceptibility relies on the parameters of the fermion system, $\mathbf{g}=\{g_1,g_2,...\}$. We assume that Eq.\eqref{eq29} is satisfied at the critical parameter $\mathbf{g}_{cri}$, and then we try to find out what is the physical consequences of such a susceptibility condition.


\subsection{1.2 Spontaneous kinetic frustration for bosons}
In the following, we focus on the case where  $\Lambda$ is a momentum loop $S^1$ with radius $Q$. Compared to $\Lambda$ being a generic manifold, this is a more common case in realistic physical systems.
Using the notations introduced above, we first write Eq.\eqref{eq24} into
\begin{equation}\label{eq31}
  S_{eff}=-\sum_{q}\chi_{ab}(q)b^{\dagger}_{q,a}b_{q,b}-\sum_qV^{-1}(\mathbf{q})b^{\dagger}_{q,a}b_{q,a}.
\end{equation}
Making the unitary transformation that diagonalizes $\chi_{ab}(q)$, i.e.,
\begin{eqnarray}
  \tilde{b}_{q,n} &=& U_{nb}(q)b_{q,b}, \\
  \tilde{b}^{\dagger}_{q,n} &=& b^{\dagger}_{q,a}U^{\dagger}_{na}(q),
\end{eqnarray}
one obtains
\begin{equation}\label{eq32}
  S_{eff}=-\sum_{q,n}\chi_{n}(q)\tilde{b}^{\dagger}_{q,n}\tilde{b}_{q,n}-\sum_{q,n}V^{-1}(\mathbf{q})\tilde{b}^{\dagger}_{q,n}\tilde{b}_{q,n},
\end{equation}
where
\begin{equation}\label{eq33}
  \chi_{n}(q)\delta_{nn^{\prime}}=U(q)\boldsymbol{\chi}(q)U^{\dagger}(q).
\end{equation}
As mentioned above,we focus on the cases where the bare Green's function $G_{0,\alpha\beta}$ is diagonal. For the cases where $G_{0,\alpha\beta}$ has non-vanishing off-diagonal entries, the following approach applies as long as the additional transformation matrix $U(q)$ is further taken into account.  In the cases $G_{0,\alpha\beta}$ is diagonal, the generalized susceptibility is cast into
\begin{equation}\label{eq34}
  \boldsymbol{\chi}(q)\equiv\chi_{ab}(q)=\chi_a(q)\delta_{ab},
\end{equation}
with $a=(\alpha,\gamma)$, i.e., a diagonal matrix. Correspondingly, the unitary transformation matrix becomes $U(q)=\mathbb{I}$. Then, Eq.\eqref{eq31} can be decomposed into the sum of independent sectors of bosons with flavor $a$, namely,
\begin{equation}\label{eq35}
  S_{eff}=\sum_{a} S^{(a)}_{eff},
\end{equation}
where
\begin{equation}\label{eq36}
  S^{(a)}_{eff}=-\sum_{q}[\chi_{a}(q)+V^{-1}(\mathbf{q})]b^{\dagger}_{q,a}b_{q,a}.
\end{equation}
For the N-flavor fermion systems, there are formally $N^2$ entries in $a$, denoting $N^2$ different ways of pairings between the fermions. Thus, Eq.\eqref{eq35}, \eqref{eq36} generally describe all possible bosons ($N^2$) that could emerge from the fermions with $N$ species.

Now we calculate $\chi_a(q)$, i.e.,
\begin{equation}\label{eq37}
  \chi_{a}(q)=-\sum_kG_{0,\alpha}(k)G_{0,\gamma}(k+q)=-\sum_{\mathbf{k},i\omega_n}\frac{1}{i\omega_n-\epsilon_{\alpha}(\mathbf{k})}\frac{1}{i\omega_n+i\nu_n-\epsilon_{\gamma}(\mathbf{k}+\mathbf{q})},
\end{equation}
where $\epsilon_{\alpha}(\mathbf{k})$ denotes the kinetic energy of the fermions with flavor $\alpha$. For non-relativistic free fermions, we assume $\epsilon(\mathbf{k})=|\mathbf{k}|^2/2m_{\alpha}$ where $m_{\alpha}$ is the effective mass of the $\alpha$-fermions.

\begin{figure*}[t]
\includegraphics[width=0.75\linewidth]{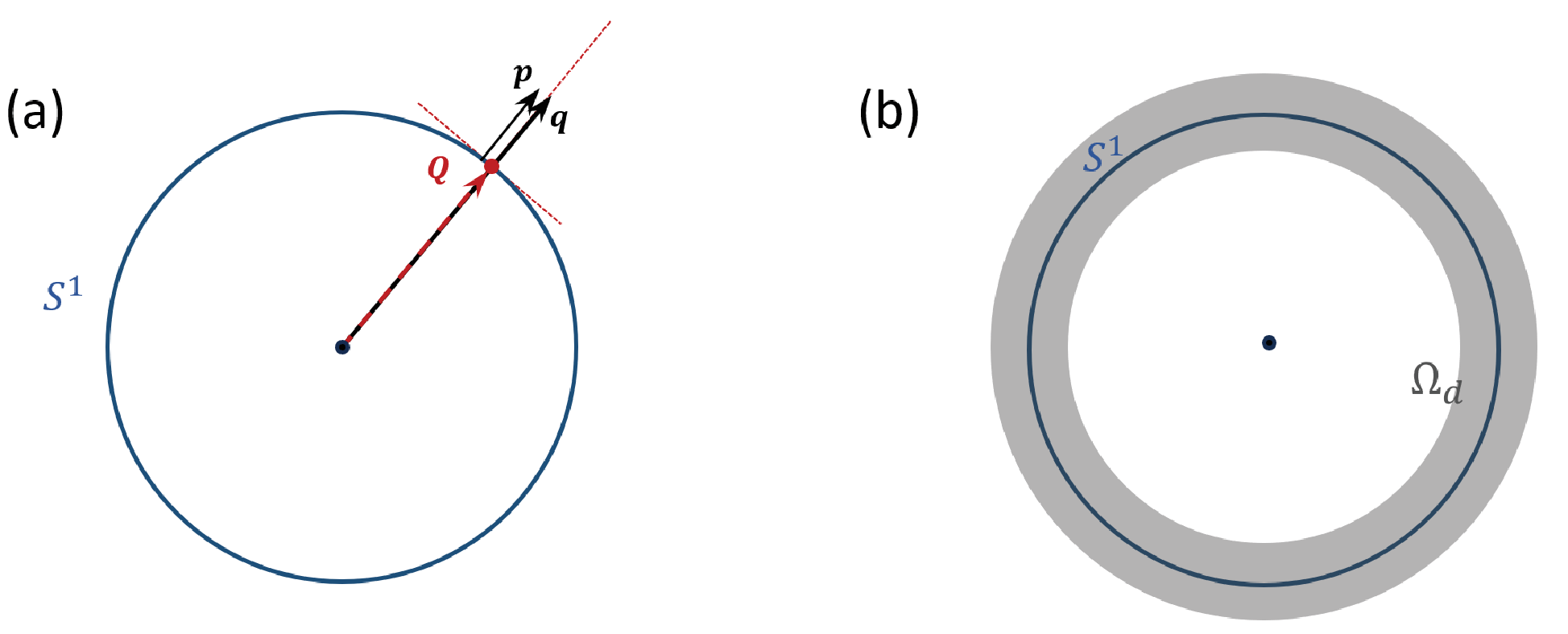}
\caption{(a) Plot of the manifold $\Lambda=S^1$. We parameterize a momentum around $\Lambda=S^1$ by a local coordinate on $\Lambda=S^1$, as denoted by the red dashed lines. $\mathbf{p}=\mathbf{q}-\mathbf{Q}$ is the relative momentum measured from $\Lambda=S^1$. (b) shows the momentum shell around $S^1$, in which the derived effective theory Eq.\eqref{eq48} well describes the low-energy physics.}\label{figadd1}
\end{figure*}

Recalling that we are working under the condition in Eq.\eqref{eq29} with $\Lambda=S^1$, which states that for $\mathbf{q}\in\Lambda$, the static part of the term in the bracket of Eq.\eqref{eq36} equals to zero. As discussed above, this means the saddle point equations are satisfied by all momentums within $\Lambda=S^1$. Thus, the low-energy bosons have momentum around $\Lambda$.
For $\Lambda=S^1$, the momentum $\mathbf{q}$ can be written in the local polar coordinate in indicated by Fig.\ref{figadd1}(a). Thus, $\mathbf{q}=\mathbf{Q}+\mathbf{p}$, where $\mathbf{p}$ is the relative momentum measured from $\mathbf{Q}$ on $S^1$. Since in the low-energy window, the long-wave fluctuations around the saddle point play the dominant role, we insert $\mathbf{q}=\mathbf{Q}+\mathbf{p}$ into Eq.\eqref{eq37} and make expansion of $\chi_a(q)$ in terms of $|\mathbf{p}|$ and $i\nu_n$.

After the sum of the Matsubara frequency in Eq.\eqref{eq37}, we obtain
\begin{equation}\label{eq38}
  \chi_{a}(i\nu_n,\mathbf{p})=-\sum_{\mathbf{k}}\frac{n_F(\epsilon_{\alpha}(\mathbf{k}))-n_F(\epsilon_{\gamma}(\mathbf{k}+\mathbf{p}+\mathbf{Q}))}{i\nu_n+\epsilon_{\alpha}(\mathbf{k})-\epsilon_{\gamma}(\mathbf{k}+\mathbf{p}+\mathbf{Q})}.
\end{equation}
Making expansion with respect to $\mathbf{p}$ and $i\nu_n$ in Eq.\eqref{eq38}, we obtain
\begin{equation}\label{eq39}
  \chi_a(i\nu_n,\mathbf{p})=-\mathcal{L}_{a,1}-\mathcal{L}_{a,2}\epsilon_{\gamma}(\mathbf{p})-\mathcal{L}_{a,3}\epsilon_{\gamma}(\mathbf{p})+i\nu_n\mathcal{L}_{a,2},
\end{equation}
where
\begin{equation}\label{eq40}
  \mathcal{L}_{a,1}=\sum_{\mathbf{k}}\frac{n_F(\epsilon_{\alpha}(\mathbf{k}))-n_F(\epsilon_{\gamma}(\mathbf{k}+\mathbf{Q}))}{\epsilon_{\alpha}(\mathbf{k})-\epsilon_{\gamma}(\mathbf{k}+\mathbf{Q})},
\end{equation}
\begin{equation}\label{eq40ad}
  \mathcal{L}_{a,2}=\sum_{\mathbf{k}}\frac{n_F(\epsilon_{\alpha}(\mathbf{k}))-n_F(\epsilon_{\gamma}(\mathbf{k}+\mathbf{Q}))}{[\epsilon_{\alpha}(\mathbf{k})-\epsilon_{\gamma}(\mathbf{k}+\mathbf{Q})]^2},
\end{equation}
\begin{equation}\label{eq41}
  \mathcal{L}_{a,3}=\sum_{\mathbf{k}}\frac{n_F(\epsilon_{\alpha}(\mathbf{k}))-n_F(\epsilon_{\gamma}(\mathbf{k}+\mathbf{Q}))}{[\epsilon_{\alpha}(\mathbf{k})-\epsilon_{\gamma}(\mathbf{k}+\mathbf{Q})]^3}\frac{2|\mathbf{k}+\mathbf{Q}|^2}{m_{\gamma}}\cos^2\theta,
\end{equation}
where $\theta$ denotes the angle between $\mathbf{p}$ and $\mathbf{k}+\mathbf{Q}$. Note that $\mathcal{L}_{a}$'s have rotational invariance after the sum of $\mathbf{k}$.
Inserting Eq.\eqref{eq39}-\eqref{eq41} to Eq.\eqref{eq36}, one obtains
\begin{equation}\label{eq44}
  S^{(a)}_{eff}=\sum_{i\nu_n,\mathbf{p}}[-i\nu_n+\frac{|\mathbf{p}|^2}{2\tilde{m}_{a}}-\tilde{\mu}_{a}]b^{\dagger}_{q,a}b_{q,a},
\end{equation}
where
\begin{equation}\label{eq45}
  \tilde{m}_{a}=\frac{\mathcal{L}_{a,2}}{\mathcal{L}_{a,2}+\mathcal{L}_{a,3}}m_{\gamma},
\end{equation}
\begin{equation}\label{eq46}
  \tilde{\mu}_a=-\frac{\mathcal{L}_{a,1}-V^{-1}(\mathbf{p})}{\mathcal{L}_{a,2}}.
\end{equation}
A key conclusion can be drawn from Eq.\eqref{eq46}. Since $\mathcal{L}_{a,1}=-\chi_a(0,\mathbf{p}=0)=-\chi_a(0,\mathbf{Q})$, $\tilde{\mu}_a=0$ is equivalent to $V^{-1}(\mathbf{q})+\chi_a(0,\mathbf{Q})=0$, i.e., $1+V(\mathbf{q})\chi_a(0,\mathbf{Q})=0$. This is just our starting condition Eq.\eqref{eq29} at $\mathbf{g}=\mathbf{g}_{cri}$. Thus, under Eq.\eqref{eq29}, $\tilde{\mu}_a=0$ is always ensured. This observation justifies the low-density condition near $\mathbf{g}_{cri}$ used in the main text.

In the above expansion over $i\nu_n$, an imaginary part proportional to the density of states of the fermions has been neglected, which arises from the imaginary part in Eq.\eqref{eq38} after taking the continuation $i\nu_n\rightarrow\nu+i0^+$. Taking this into account, an additonal imaginary term takes place in Eq.\eqref{eq46}, giving rise to
\begin{equation}\label{eq47}
  S^{(a)}_{eff}=\sum_{i\nu_n,\mathbf{p}}[-i\nu_n+\frac{|\mathbf{p}|^2}{2\tilde{m}_{a}}-\tilde{\mu}_{a}-i\gamma_a]b^{\dagger}_{q,a}b_{q,a}.
\end{equation}
Eq.\eqref{eq47} constitutes the action describing the $a$-flavor bosons generated from correlated fermions. The $\gamma$ term originates from the renormalization of bosons from its fermionic environment, resulting in the finite lifetime of bosons.

Recalling that $\mathbf{p}=\mathbf{q}-\mathbf{Q}$ and $\mathbf{q}$ is in parallel with $\mathbf{Q}$, we obtain the action of the bosons as $S_{eff}=\sum_a S^{(a)}_{eff}$, and
\begin{equation}\label{eq48}
  S^{(a)}_{eff}=\sum_{i\nu_n,\mathbf{q}\in\Omega_d}[-i\nu_n+\frac{(|\mathbf{q}|-Q)^2}{2\tilde{m}_{a}}-\tilde{\mu}_{a}-i\gamma_a]b^{\dagger}_{q,a}b_{q,a}.
\end{equation}
A momentum cutoff has to be introduced since Eq.\eqref{eq48} only describes the low-energy physics, thus the sum of $\mathbf{q}$ is restricted to a thin momentum shell $\Omega_d$ around the manifold $\Lambda=S^1$, as shown by Fig.\ref{figadd1}(b).

\subsection{1.3 Correlated bosons on the manifold}
Due to the degenerate kinetic energy along $S^1$ in Eq.\eqref{eq48}, the boson-boson interaction could play the dominant role. Thus, the higher order fluctuations effect, which induces the boson-boson interaction, must be taken into account. The next leading order expansion is $\mathrm{Tr}(G_0\Sigma)^4/4$ in Eq.\eqref{eq21}. This term leads to
\begin{equation}\label{eq49}
  S_{I,eff}=4\sum_{k,q_1,q_2,q_3}G_{0,\alpha\beta}(k)G_{0\delta\gamma}(k+q_1)G_{0,\rho\mu}(k+q_1-q_2)G_{0\nu\lambda}(k+q_1-q_2+q_3)b^{\dagger}_{q_1,\beta\delta}b_{q_1-q_2+q_3,\lambda\alpha}b^{\dagger}_{q_3,\mu\nu}b_{q_2,\gamma\rho}.
\end{equation}
In the case $G_{0,\alpha\beta}$ is diagonal, the above term can be decomposed into $S_{I,eff}=S^{intra}_{I,eff}+S^{inter}_{I,eff}$, where $S^{intra}_{I,eff}$ is the intra-flavor interaction and $S^{inter}_{I,eff}$ the inter-flavor interaction. $S^{intra}_{I,eff}$ is given by
\begin{equation}\label{eq50}
  S^{intra}_{I,eff}=\sum_{q_1,q_2,q_3}U_a(q_1,q_2,q_3)b^{\dagger}_{q_1,a}b_{q_1-q_2+q_3,a}b^{\dagger}_{q_3,a}b_{q_2,a},
\end{equation}
where $a=(\alpha,\delta)$, and we defined
\begin{equation}\label{eq50a}
  U_a(q_1,q_2,q_3)=4\sum_{k}G_{0,\alpha}(k)G_{0,\delta}(k+q_1)G_{0,\alpha}(k+q_1-q_2)G_{0,\delta}(k+q_1-q_2+q_3).
\end{equation}
Besides, the inter-flavor interaction has the general form
\begin{equation}\label{eq51}
  S^{inter}_{I,eff}=\sum_{q_1,q_2,q_3}U_{abcd}(q_1,q_2,q_3)b^{\dagger}_{q_1,a}b_{q_1-q_2+q_3,b}b^{\dagger}_{q_3,c}b_{q_2,d},
\end{equation}
where bosons with different flavors couple to each other.

It should be noted that the kinetic frustration in Eq.\eqref{eq48} is only active in the Hilbert space of bosons with $a$ flavor. Since the infinite degeneracy lies in each $a$-flavor subspace, it is the intra-flavor, rather than the inter-flavor, interaction $S^{(4)}_{b,intra}$ that becomes dominant. In this sense, novel physics, if any, could only arise from $S^{intra}_{I,eff}$ instead of $S^{inter}_{I,eff}$. Moreover, for most cases, once bosons are formed in a certain channel, this leading instability would suppress bosons in other channels. Thus, only one single favor needs to be considered in the general realistic cases. Hence, the formally derived inter-flavor interaction is absent in these cases.  For example, for the $N=2$ fermion system with the inter-flavor fermion paring, i.e, $b_{q,\alpha\beta}=b_{q,\alpha\overline{\alpha}}$, the inter-flavor boson-boson interaction is clearly absent.

The interaction vertex in Eq.\eqref{eq50} in general has $q$-dependence. Following the standard way of deriving the Ginzberg-Landau theory, the vertex can be further simplified in the long-wave approximation, since the higher order $q$-dependent terms in the Taylor expansion of the vertex generally have the smaller scaling dimensions, and are more irrelevant in renormalization group sense. Thus, Eq.\eqref{eq50} and Eq.\eqref{eq50a} are reduced to
\begin{equation}\label{eq52}
  S^{(a)}_{I,eff}=U_a\sum_{q_1,q_2,q_3}b^{\dagger}_{q_1,a}b_{q_1-q_2+q_3,a}b^{\dagger}_{q_3,a}b_{q_2,a},
\end{equation}
where $a=(\alpha,\delta)$, and we defined
\begin{equation}\label{eq53}
  U_a=4\sum_{k}G_{0,\alpha}(k)G_{0,\delta}(k)G_{0,\alpha}(k)G_{0,\delta}(k).
\end{equation}

Then, collecting both Eq.\eqref{eq48} and Eq.\eqref{eq52}, the dominant low-energy physics is described by $S_{eff}=\sum_a S^{(a)}_{eff}$, where
\begin{equation}\label{eq54}
\begin{split}
  S^{(a)}_{eff}&=\sum_{i\nu_n,\mathbf{q}\in\Omega_d}[-i\nu_n+\frac{(|\mathbf{q}|-Q)^2}{2\tilde{m}_{a}}-\tilde{\mu}_{a}-i\gamma_a]b^{\dagger}_{q,a}b_{q,a}+U_a\sum_{q_1,q_2,q_3}b^{\dagger}_{q_1,a}b_{q_1-q_2+q_3,a}b^{\dagger}_{q_3,a}b_{q_2,a}.
\end{split}
\end{equation}
Eq.\eqref{eq54} describes correlated bosons emergent from interacting fermions. Interestingly, the bosons have their kinetic energy minimized along the 1D manifold $\Lambda=S^1$. Note that the above derivations and thus Eq.\eqref{eq51} also apply for the case where $\Lambda$ is a generic 1D manifold. In the above derivations, no concrete models are specified. Besides, except for the susceptibility condition in Eq.\eqref{eq29}, no other conditions are required.  Therefore, the emergent bosons with kinetic frustration is a general result of Eq.\eqref{eq29}.

Here,  the higher-order interactions involving six-bosons or eight-bosons are neglected. This is justified by the fact that higher-order interaction terms are more irrelevant in the renormalization group sense, as can be understood by the following analysis.  The key feature of the kinetic term is the flatness along the loop $S^1$. At low-energy, the bosons are restricted to the phase space along the loop, mimicking 1D physics with flat band described by $H_{1D}=\int dq\epsilon_0b^{\dagger}_qb_{q}$, for which the dimension of the boson fields is 1/2. Thus, the fourth-order interaction term (restricted to the low-energy phase space) reads as $U\int dq_1dq_2dq_3/(2\pi)^3b^{\dagger}_{q_1}b^{\dagger}_{q_2}b_{q_3}b_{q_1+q_2-q_3}$, which has the scaling dimension, $-1$. This suggests that the long-range orders are disfavored (but the possible development of topological orders cannot be excluded). The six-order interaction term $U_6\int dq_1dq_2dq_3dq_4dq_5/(2\pi)^3b^{\dagger}_{q_1}b^{\dagger}_{q_2}b^{\dagger}_{q_3}b_{q_4}b_{q_5}b_{q_1+q_2+q_3-q_4-q_5}$ has the dimension $-2$, suggesting that it is more irrelevant in the renormalization group sense. Hence, in Eq.\eqref{eq54}, it is sufficient to keep the leading order interaction term.

\section{2. Formation of topological orders with composite fermions}
From Eq.\eqref{eq54}, we can read off the effective Hamiltonian for the emergent bosons with flavor $a$, namely, $H_{eff}=H_K+H_{U}$. $H_K$ describes the kinetic energy of bosons
\begin{equation}\label{eq55}
  H_K=\sum_{\mathbf{q}\in\Omega_d}[\frac{(|\mathbf{q}|-Q)^2}{2\tilde{m}}-\tilde{\mu}^{\prime}]b^{\dagger}_{\mathbf{q}}b_{\mathbf{q}},
\end{equation}
where we have omitted the flavor index $a$, since in most cases only a single flavor of bosons is relevant. $\tilde{\mu}^{\prime}=\tilde{\mu}-i\gamma$ is an effective chemical potential, whose imaginary part describes the lifetime of the bosons. The interaction $H_U$ can be written in real space as
\begin{equation}\label{eq56}
  H_U=U\sum_{\mathbf{r}}b^{\dagger}_{\mathbf{r}}b_{\mathbf{r}}b^{\dagger}_{\mathbf{r}}b_{\mathbf{r}},
\end{equation}
which is a local boson-boson interaction. The condition $\mathbf{q}\in\Omega_d$ can be loosen in the above continuum model Eq.\eqref{eq55}, which does not affect the low-energy physics around the momentum loop $S^1$. In the following, we regard $\tilde{m}$, $\tilde{\mu}^{\prime}$ and $U$ as parameters, and justify that topological orders can emerge as possible ground state, which enjoys lower energy than all boson condensates proposed to date.

It is clear that the $U$ term in Eq.\eqref{eq56} plays a crucial role. Therefore, the system would like to minimize the energy cost from the $U$ term. The bosons can greatly lower their energy by behaving as fermions, because the fermions with the same flavor cannot occupy the same spatial coordinate $\mathbf{r}$ due to the anstisymmetric nature of their wave function. Thus, we introduce the fermionic representation for the bosons, i.e., we represent the many-body bosonic wave function by antisymmetric (fermionic) many-body wave functions attached with a nonlocal $\mathrm{U}(1)$ Chern-Simons factor, i.e.,
\begin{equation} \label{eq57}
\Psi_b(\mathbf{r}_1,...,\mathbf{r}_N)=\Psi_f(\mathbf{r}_1,...,\mathbf{r}_N)e^{i\sum_{i<j}\mathrm{arg}[\mathbf{r}_i-\mathbf{r}_j]}.
\end{equation}
and
\begin{equation} \label{eq58}
e^{i\sum^N_{i<j}\mathrm{arg}[\mathbf{r}_i-\mathbf{r}_j]}=\prod^N_{i<j}\frac{z_i-z_j}{|z_i-z_j|},
\end{equation}
where we introduced $z_{i}=x_{i}+iy_i$ to denote the coordinate for the fermion at $\mathbf{r}_i$. The CS phase in Eq.\eqref{eq58} reproduces the bosonic statistics from fermions. The advantage of the representation is that the fermionic wave function can further lower the system energy although the CS phase has additional cost in the kinetic energy.

To calculate the cost of kinetic energy, one needs to insert the CS phase into the kinetic Hamiltonian. Then, the derivative arises as  $\nabla_l\sum^N_{i<j}\mathrm{arg}[\mathbf{r}_i-\mathbf{r}_j]$, where $\nabla_l$ denotes the derivative acting on the coordinates of the $l$-th particle. This generates the gauge field attached to the fermions, i.e.,
\begin{equation} \label{eq59}
\mathcal{A}(\mathbf{r}_l)=\nabla_l\sum^N_{i<j}\mathrm{arg}(\mathbf{r}_i-\mathbf{r}_j)=\sum^N_{j\neq l}\hat{z}\times\frac{\mathbf{r}_l-\mathbf{r}_j}{|\mathbf{r}_l-\mathbf{r}_j|^2},
\end{equation}
Thus, the $\alpha$ component of $\mathcal{A}(\mathbf{r}_l)$ reads as
\begin{equation} \label{eq60}
\mathcal{A}^{\alpha}(\mathbf{r}_l)=\epsilon^{\alpha\beta}\sum_{j\neq l}\frac{(\mathbf{r}_l-\mathbf{r}_j)_{\beta}}{|\mathbf{r}_l-\mathbf{r}_j|^2}.
\end{equation}
The gauge field generates an effective CS magnetic field, $B_{CS}(\mathbf{r}_l)=\nabla_l\times\mathcal{A}(\mathbf{r}_l)=2\pi\sum_{j\neq l}\delta(\mathbf{r}_j-\mathbf{r}_l)=2\pi n(\mathbf{r}_l)$.

Similar to the FQH systems, the smearing flux approximation is often used which treats $B_{CS}$ as uniform magnetic field. Because the Chern-Simons flux is proportional to the fermion density, this approximation is essentially a restriction to ansatz states with uniform, translational invariant densities. This is expected, since we are looking for the possible disordered ground state with topologically orders.

The uniform flux attached to the fermions leads to Landau quantization. We introduce the ladder operators as
 \begin{equation} \label{eq61}
a=\sqrt{\frac{c}{2e\hbar B_{CS}}}(\Pi_x+i\Pi_y),
\end{equation}
and
 \begin{equation} \label{eq62}
a^{\dagger}=\sqrt{\frac{c}{2e\hbar B_{CS}}}(\Pi_x-i\Pi_y),
\end{equation}
where $\Pi_x=q_x-\frac{e}{c}A_x$ and $\Pi_y=q_y-\frac{e}{c}A_y$. Then, the kinetic energy of the fermions in Eq.\eqref{eq55} is cast in the first quantized form as
 \begin{equation} \label{eq63}
H_K=\frac{1}{2\tilde{m}}[\Pi^2_x+\Pi^2_y+Q^2-2Q(\Pi^2_x+\Pi^2_y)^{1/2}]-\tilde{\mu}^{\prime}.
\end{equation}
Using $\Pi^2_x+\Pi^2_y=\frac{2e\hbar B_{CS}}{c}(a^{\dagger}a+1/2)$, we obtain the energy of the Landau levels, i.e.,
 \begin{equation} \label{eq64}
E_l=\frac{Q^2}{2\tilde{m}}[\sqrt{(l+\frac{1}{2})\frac{\omega_c}{Q^2/2\tilde{m}}}-1]^2-\tilde{\mu}^{\prime},
\end{equation}
where $\omega_c=B_{CS}/m_b$ and we have taken $e=\hbar=c=1$. Clearly, the energy minima are located at $B_{CS}=\frac{Q^2}{2(l+1/2)}$. Since $B_{CS}=2\pi n$, where $n$ is the average density of bosons, The energy is minimized as long as the following condition is met, i.e.,
\begin{equation} \label{eq65}
n_l=\frac{Q^2}{4\pi(l+1/2)}.
\end{equation}
This is the optimal density that exhibits the lowest energy in the language of fermions, which corresponds to the fully filled lowest Landau level.

From above, we know that in the language of composite fermions, the ground state wave function is the fully filled Landau level. We use $\Psi^l_f(\mathbf{r}_1,...,\mathbf{r}_N)$ to denote the fermionic wave function of the $l$-th fully filled Landau level, which is given by
  \begin{equation} \label{eqn106}
\Psi^l_f(\mathbf{r}_1,...,\mathbf{r}_N)=\frac{1}{\sqrt{N!}}\mathrm{det}_{m,j}[\chi^l_m(z_j)]
\end{equation}
where $\chi^l_m(z_j)$ can be obtained by solving the eigenvectors of $H_0$ under symmetric gauge, and it is obtained as,
  \begin{equation} \label{eqn107}
\chi^l_m(z_j)=\frac{(-1)^l\sqrt{l!}}{l_B\sqrt{2\pi 2^m(l+m)!}}(\frac{z}{l_B})^me^{-\frac{|z|^2}{4l^2_B}}L^{(m)}_l[\frac{|z|^2}{2l^2_B}],
\end{equation}
where $m=-l,...,N-l$ and $L^m_l(x)$ is the adjoint Laguerre polynomial, and the magnetic length $l_B=1/\sqrt{2\pi n}$. From Eq. \eqref{eq57}, we note that the composite fermion here is essentially a boson attached to 1-flux quanta.

Then, the ground state energy of the composite fermion state can be obtained by  calculating the expectation of $H_K$ with respect to the boson wave function. This leads to
\begin{equation}\label{eq66}
E_{TO}=\langle \Psi_b |[\frac{(|\mathbf{q}|-Q)^2}{2\tilde{m}}]|\Psi_b\rangle=\frac{\pi^2 n^2}{2\tilde{m}Q^2}\mathrm{log}^2\frac{4n}{Q^2},
\end{equation}
Therefore,  $E_{TO}\propto n^2\log^2n$. For relatively low density, this is lower than the energy of conventional boson condensation (FF or LO state) with $E_{FF/LO}\propto n$. It is also lower than the energy of the non-uniform fragmented condensation state with $E_{FC}\propto n^{4/3}$ \cite{Gopalakrishnans}. A more detailed comparison is shown by Fig.\ref{figadd2} below.

Here, we note that the ground state energy in Eq.\eqref{eq66} is independent of $U$, which seems a bit confusing. This is because of the fact that the ground state lies in a Hilbert subspace where each local position $\mathbf{r}$ is either empty or singly-occupied. Due to the strong kinetic frustration for bosons, the interaction $U$ plays the dominant role, and the bosons are subject to strong on-site repulsion, thus they behave as hardcore bosons. Therefore, only the empty or singly-occupied subspace is favored, in which the interaction term U is effectively projected out.

In the remaining sections, we will discuss application examples of the above general theory. We first predict novel excitonic topological orders formed out of strongly-correlated doped semiconductors and then discuss chiral spin liquids formed in frustrated quantum magnets. The general theory unifies two completely different topological orders within the same framework and reveals their common underlying mechanism.

\section{3. Excitonic topological order formed out of strongly-correlated doped semiconductors}

\subsection{3.1 Excitonic moat band from doped semiconductors}
We start from a two band model, $H=H_0+H_I$, where the free Hamiltonian $H_K$ is given by
\begin{equation}\label{eqad1}
  H_0=\sum_{\mathbf{k},n,\sigma}\epsilon_n(\mathbf{k})c^{\dagger}_{n,\mathbf{k},\sigma}c_{n,\mathbf{k},\sigma},
\end{equation}
where $n$ and $\sigma$ represent for the band and spin of fermions, respectively. $\epsilon_{\pm}(\mathbf{k})=\pm k^2/2m\pm D/2-\mu_0$, where $D>0$ is the band gap and $\mu_0$ is the chemical potential determined by the level of doping.

In the following, we are interested in the doped semiconductor case where only the electron (or hole) Fermi surface arises, as shown in Fig.2(a) of the main text, which requires $0<D<2|\mu_0|$. Then, we consider the short-range inter-band interaction between the fermions, i.e.,
\begin{equation}\label{eqad2}
  H_I=V\sum_{\mathbf{k},\mathbf{k}^{\prime},\mathbf{q}}\sum_{\sigma,\sigma^{\prime}}c^{\dagger}_{+,\mathbf{k}+\mathbf{q},\sigma}c_{+,\mathbf{k},\sigma}c^{\dagger}_{-,\mathbf{k}^{\prime}-\mathbf{q},\sigma^{\prime}}c_{-,\mathbf{k}^{\prime},\sigma^{\prime}}.
\end{equation}

As the first step, we calculate the dispersion of excitons.  The intra-band interactions can be neglected as they only affect the dispersion of excitons via the mass renormalization. We introduce the exciton operator with total momentum $\mathbf{p}$ that annihilates an electron in the valence band and creates an electron in the conduction band, i.e., $\Delta^{\dagger}_{\mathbf{p}}=\sum_{\mathbf{k}}\phi_{\mathbf{p}}(\mathbf{k})c^{\dagger}_{+,\mathbf{k},\sigma}c_{-,\mathbf{p}-\mathbf{k},\sigma}$, where $\phi_{\mathbf{p}}(\mathbf{k})$ is a variational parameter. The sum of $\mathbf{k}$ only involves momenta larger than the Fermi momentum $|\mathbf{k}|>k_F$ due to Pauli's exclusion. We then calculate and minimize the expectation value $W_{ex}=\langle FS|\Delta_{\mathbf{p}}[H-E_{ex}(\mathbf{p})] \Delta^{\dagger}_{\mathbf{p}}|FS\rangle$, where $E_{ex}(\mathbf{p})$ is a Lagrangian multiplier that normalizes the exciton states and $|FS\rangle$ is the Fermi sea state with the electron (or hole) Fermi surface. The minimization of $W_{ex}$ leads to the variational equation, $\delta W_{ex}/\delta\phi_{\mathbf{p}}(\mathbf{k})=0$, which produces the energy dispersion of excitons as excitations on top of the ground state $|FS\rangle$.

Calculating $W_{ex}$ and taking its variation with respect to $\phi_{\mathbf{p}}(\mathbf{k})$ leads to the following equation:
\begin{equation}\label{eqad3}
  [\epsilon_+(\mathbf{k})-\epsilon_-(\mathbf{k}-\mathbf{p})-E_{ex}(\mathbf{p})]\phi^{\star}_{\mathbf{p}}(\mathbf{k})=V\sum_{|\mathbf{k}^{\prime}|>k_F}\phi^{\star}_{\mathbf{p}}(\mathbf{k}^{\prime}).
\end{equation}
We define
\begin{equation}\label{eqad4}
  A_{\mathbf{p}}=\sum_{|\mathbf{k}^{\prime}|>k_F}\phi^{\star}_{\mathbf{p}}(\mathbf{k}^{\prime}),
\end{equation}
then Eq.\eqref{eqad3} and Eq.\eqref{eqad4} generate the self-consistent equation,
\begin{equation}\label{eqad5}
  1=V\sum_{|\mathbf{k}|>k_F}\frac{1}{\epsilon_+(\mathbf{k})-\epsilon_-(\mathbf{k}-\mathbf{p})-E_{ex}(\mathbf{p})}
\end{equation}
This further leads to
\begin{equation}\label{eqad6}
  1=V\rho_0\int^1_{\mu^{\prime}_0-D^{\prime}/2}d\epsilon^{\prime}\frac{1}{2\epsilon^{\prime}+\epsilon^{\prime}_p-2\sqrt{2\epsilon^{\prime}_p\epsilon^{\prime}}+D^{\prime}-E^{\prime}_{ex}(\mathbf{p})},
\end{equation}
where $\rho_0$ the density of states of the 2D electrons. $\epsilon^{\prime}=\epsilon/\Lambda$, $\epsilon^{\prime}_p=\epsilon_p/\Lambda$ with $\epsilon_p=p^2/2m$, $D^{\prime}=D/\Lambda$ and $E^{\prime}_{ex}=E_{ex}/\Lambda$, which are dimensionless quantities normalized by the energy cutoff $\Lambda$. The exciton energy dispersion, i.e., $E^{\prime}_{ex}(\mathbf{p})$, can be solved from Eq.\eqref{eqad6}. As shown by Fig.2(c),(d) of the main text,  excitonic moat band is found to occur for the finite-doping case with a metallic Fermi surface.

\subsection{3.2 Mean-field phase diagram}
We now neglect for a time the effect of quantum fluctuations, and study the model $H_0+H_I$ in the mean-field level. We let $\tilde{\mathbf{q}}=\mathbf{k}+\mathbf{q}-\mathbf{k}^{\prime}$, the interaction term $H_I$ is written into:
\begin{equation}\label{eqad7}
H_I=-V\sum_{\mathbf{k},\mathbf{p},\tilde{\mathbf{q}}}\sum_{\sigma,\sigma^{\prime}}c^{\dagger}_{+,\mathbf{p},\sigma}c_{-,\mathbf{p}-\tilde{\mathbf{q}},\sigma^{\prime}}c^{\dagger}_{-,\mathbf{k},\sigma^{\prime}}c_{+,\mathbf{k}+\tilde{\mathbf{q}},\sigma},
\end{equation}
where we have introduced $\mathbf{k}^{\prime}\rightarrow\mathbf{k}^{\prime}-\tilde{\mathbf{q}}$, $\mathbf{k}\rightarrow\mathbf{k}+\tilde{\mathbf{q}}$ and then replaced $\mathbf{k}^{\prime}$ by $\mathbf{p}$ for brevity.

We then introduce the excitonic order parameters as $\Delta_{\tilde{\mathbf{q}}}=V\sum_{\mathbf{p}}\langle c^{\dagger}_{+,\mathbf{p},\sigma}c_{-,\mathbf{p}-\tilde{\mathbf{q}},\sigma}\rangle$ and $\Delta^{\star}_{\tilde{\mathbf{q}}}=V\sum_{\mathbf{p}}\langle c^{\dagger}_{-,\mathbf{p},\sigma}c_{+,\mathbf{p}+\tilde{\mathbf{q}},\sigma}\rangle$ to decouple the interaction. This leads to
\begin{equation}\label{eqad8}
H_I=-\sum_{\mathbf{k},\tilde{\mathbf{q}},\sigma}[\Delta_{\tilde{\mathbf{q}}}c^{\dagger}_{-,\mathbf{k},\sigma}c_{+,\mathbf{k}+\tilde{\mathbf{q}},\sigma}+h.c.]+2\sum_{\tilde{\mathbf{q}}}\Delta^2_{\tilde{\mathbf{q}}}/V,
\end{equation}
The factor of 2 in the last term is due to the sum of the spin. For a ground state with $\Delta_{\tilde{\mathbf{q}}}\neq0$, the inter-band particle-hole excitations are energetically favored, leading to condensation of the electron-hole pairs with momentum $\tilde{\mathbf{q}}$ and total spin zero.  The mean-field Hamiltonian can be written in the basis $\psi_{\mathbf{k},\mathbf{q}}=[c_{+,\mathbf{k}+\mathbf{q},\sigma},c_{-,\mathbf{k},\sigma}]^T$ as,
\begin{equation}\label{eqad9}
H_{MF}=\sum_{\mathbf{k}\sigma}
\left(
  \begin{array}{ccc}
    c_{+,\mathbf{k}+\mathbf{q},\sigma}\\
    c_{-,\mathbf{k},\sigma}\\
  \end{array}
\right)^{\dagger}
\left(
  \begin{array}{ccc}
    \epsilon_{c,\mathbf{k}+\mathbf{q}}-\mu_0+D/2 & -\Delta^{\star}_{\mathbf{q}} \\
    -\Delta_{\mathbf{q}} & \epsilon_{v,\mathbf{k}}-\mu_0-D/2 \\
  \end{array}
\right)
\left(
  \begin{array}{ccc}
    c_{+,\mathbf{k}+\mathbf{q},\sigma} \\
    c_{-,\mathbf{k},\sigma} \\
  \end{array}
\right)   +\frac{2|\Delta_{\mathbf{q}}|^2}{V},
\end{equation}
where $\epsilon_{c,\mathbf{k}}=k^2/2m$, $\epsilon_{v,\mathbf{k}}=-k^2/2m$. We further define
\begin{equation}\label{eqad10}
\epsilon_{1,\mathbf{k},\mathbf{q}}=[\epsilon_{c,\mathbf{k}+\mathbf{q}}+\epsilon_{v,\mathbf{k}}]/2-\mu_0
\end{equation}
and
\begin{equation} \label{eqad11}
\epsilon_{2,\mathbf{k},\mathbf{q}}=[\epsilon_{c,\mathbf{k}+\mathbf{q}}-\epsilon_{v,\mathbf{k}}]/2+D/2.
\end{equation}
Then, the mean-field Hamiltonian is written as,
\begin{equation}\label{eqad12}
H_{MF}=\sum_{\mathbf{k},\sigma}\psi^{\dagger}_{\mathbf{k},\mathbf{q}}
\left(
  \begin{array}{ccc}
   \epsilon_{1,\mathbf{k},\mathbf{q}}+\epsilon_{2,\mathbf{k},\mathbf{q}} & -\Delta_{\mathbf{q}} \\
    -\Delta_{\mathbf{q}} & \epsilon_{1,\mathbf{k},\mathbf{q}}-\epsilon_{2,\mathbf{k},\mathbf{q}} \\
  \end{array}
\right)   \psi_{\mathbf{k},\mathbf{q}}+\frac{2|\Delta_{\mathbf{q}}|^2}{V}.
\end{equation}
The energy dispersion is then readily obtained as
\begin{equation}\label{eqad13}
E_{\pm,\mathbf{k},\mathbf{q}}=\pm\sqrt{\epsilon^2_{2,\mathbf{k},\mathbf{q}}+\Delta^2_{\mathbf{q}}}+\epsilon_{1,\mathbf{k},\mathbf{q}}.
\end{equation}
By minimizing the free energy derived from Eq.\eqref{eqad13}, we obtain the self-consistent equation for the excitonic order parameter, i.e.,
\begin{equation}\label{eqad16}
\Delta_{\mathbf{q}}=-\frac{V}{2N_s}\sum_{\mathbf{k},l=\pm}l\cdot n_F(E_{l,\mathbf{k},\mathbf{q}}-\mu)\frac{\Delta_{\mathbf{q}}}{\sqrt{\epsilon^2_{2,\mathbf{k},\mathbf{q}}+\Delta^2_{\mathbf{q}}}}.
\end{equation}
One can self-consistently solve the above equation with respect to the order parameter $\Delta_{\mathbf{q}}$. The phase diagram with varying $D$ and $\mu_0$ has been shown in Fig.2(b) of the main text. With lowering $|\mu_0|$ down to a critical value $\mu_{0,cri}$, exciton condensation at finite momentums takes place, leading to the FF/LO excitonic insulator. With further lowering $|\mu_0|$, condensation at zero momentum is then favored, resulting in a zero-momentum BCS excitonic insulator.

\subsection{3.3 Excitonic topological order beyond the mean-field level}
Since the excitons formed can exhibit moat band, as shown in Fig.2(d) of the main text, the effect of quantum fluctuations could be essential. Thus, fluctuations beyond the mean-field level should be carefully analyzed. This can be achieved using the general method introduced by Sec.1 and Sec.2.

We first made Hubbard-Stratonovich decomposition of the interaction $H_I$ by introducing the bosonic fields,
\begin{eqnarray}
  \Delta_{\mathbf{q}} &=& V\sum_{\mathbf{p}}c^{\dagger}_{+,\mathbf{p},\sigma}c_{-,\mathbf{p}-\mathbf{q},\sigma}, \\
  \Delta^{\dagger}_{\mathbf{q}} &=& V\sum_{\mathbf{p}}c^{\dagger}_{-,-\mathbf{p}-\mathbf{q},\sigma}c_{+,\mathbf{p},\sigma}.
\end{eqnarray}
The bosonic fields can be understood as electron-hole pairs, i.e., favored by the inter-band interaction $V$. The partition function describing the interacting electron-hole model $H_0+H_I$ is written as $Z=\int D\overline{\psi}_{\mathbf{k},\mathbf{q}}D\psi_{\mathbf{k},\mathbf{q}}D\Delta^{\dagger}_{\mathbf{q}}D\Delta_{\mathbf{q}}e^{-S}$, where
\begin{equation}\label{eqad17}
  -S=\sum_{\mathbf{k},\mathbf{q},\sigma}\int^{\beta}_0 d\tau\overline{\psi}_{\mathbf{k},\mathbf{q}}G^{-1}(\mathbf{k},\mathbf{q},\tau)\psi_{\mathbf{k},\mathbf{q}}-\frac{2}{V}\int^{\beta}_0 d\tau\sum_{\mathbf{q}}\Delta^{\dagger}_{\mathbf{q}}\Delta_{\mathbf{q}},
\end{equation}
and the imaginary-time Green's function $G^{-1}(\mathbf{k},\mathbf{q},\tau)=G^{-1}_0(\mathbf{k},\mathbf{q},\tau)-\Sigma_{\mathbf{q}}$. Here, $G^{-1}_0(\mathbf{k},\mathbf{q},\tau)$ is the inverse of the Green's function of the free electrons and holes, i.e.,
\begin{equation}\label{eqad18}
  G^{-1}_0(\mathbf{k},\mathbf{q},\tau)=\left(
                                         \begin{array}{cc}
                                           \partial_{\tau}-\epsilon_{1,\mathbf{k},\mathbf{q}} & 0 \\
                                           0 & \partial_{\tau}-\epsilon_{1,\mathbf{k},\mathbf{q}}+\epsilon_{2,\mathbf{k},\mathbf{q}} \\
                                         \end{array}
                                       \right),
\end{equation}
and $\Sigma_{\mathbf{q}}$ reads as,
\begin{equation}\label{eqad19}
  \Sigma_{\mathbf{q}}=\left(
                        \begin{array}{cc}
                          0 & -\Delta_{\mathbf{q}} \\
                          -\Delta^{\dagger}_{\mathbf{q}} & 0 \\
                        \end{array}
                      \right).
\end{equation}
Integrating out the fermionic Grassmann field in Eq.\eqref{eqad17}, we obtain the effective action that only contains the bosonic degrees of freedom, i.e.,
\begin{equation}\label{eqad20}
  -S_{eff}=\mathrm{Trln}(-G^{-1})+\frac{2}{V}\int^{\beta}_0 d\tau\sum_{\mathbf{q}}\Delta^{\dagger}_{\mathbf{q}}\Delta_{\mathbf{q}},
\end{equation}
where $\mathrm{Tr}(...)$ denotes the matrix trace and the sum of momentums. As discussed in Sec.1, we can derive the saddle point equation from Eq.\eqref{eqad20} by taking the variation $\delta S_{eff}/\delta \Delta_{\mathbf{q}}=0$, leading to $\chi(\mathbf{q},0)-V^{-1}=0$. This is exactly reduced to the mean-field equation at the critical point, e.g. $\mu_0=\mu_{0,cri}$ denoted by the red thick curve in Fig.2(f) of the main text.

Treating $\Sigma_{\mathbf{q}}$ pertabatively, which works well for the critical regime, and making expansion of the logarithmic function, one obtains the action at the Gaussian level:
  \begin{equation}\label{eqad20a}
  \begin{split}
    -S^{(2)}_{eff}&=-\mathrm{Tr}(G_0\Sigma G_0\Sigma)-\frac{2}{V}\int^{\beta}_0 d\tau\sum_{\mathbf{q}}\Delta^{\dagger}_{\mathbf{q}}\Delta_{\mathbf{q}}\\
    &=2\sum_{\mathbf{q},i\nu_n}[\chi(\mathbf{q},i\nu_n)-\frac{1}{V}]\Delta^{\dagger}(\mathbf{q},i\nu_n)\Delta(\mathbf{q},i\nu_n),
    \end{split}
  \end{equation}
where $\chi(\mathbf{q},i\nu_n)$ is the susceptibility in the particle-hole channel, i.e.,
\begin{equation}\label{eqad21}
  \chi(\mathbf{q},i\nu_n)=-\sum_{\mathbf{k},i\omega_n}\frac{1}{i\omega_n-\epsilon_{1,\mathbf{k},\mathbf{q}}-\epsilon_{2,\mathbf{k},\mathbf{q}}}\cdot\frac{1}{i\omega_n+i\nu_n-\epsilon_{1,\mathbf{k},\mathbf{q}}+\epsilon_{2,\mathbf{k},\mathbf{q}}}.
\end{equation}
Eq.\eqref{eqad20} and Eq.\eqref{eqad21} are in correspondence to Eq.\eqref{eq36} and Eq.\eqref{eq37} in the general formalism. Note that unlike the general theory for $V<0$, here the inter-band electron-electron interaction is repulsive $V>0$ , and a minus sign occurs in front of $1/V$.

For the doped semiconductor case, $0<D<2|\mu_0|$, we sum over the Matsubara frequency $i\omega_n$ in Eq.\eqref{eqad21}, and then numerically calculate the static susceptibility $\chi(\mathbf{q},0)$. In doing so, we divide the momentum space into a $\mathbf{k}$-lattice in the polar coordinate, where $k=0,dk,2dk,...$ and $\theta=0,d\theta,2d\theta,...,2\pi$.  The calculated $\chi(\mathbf{q},0)$ is shown in Fig.2(e) of the main text. As shown, the maxima are found to be degenerate along the momentum loop $S^1$ with the radius $Q$. Thus, at $\mu_0=\mu_{0,cri}$, $\chi(\mathbf{q},0)-1/V=0$ is satisfied simultaneously by all $\mathbf{q}\in S^1$, indicating an equal tendency for exciton condensation along the loop. This brings about strong frustration effect for the excitons.

The frustration enhances quantum fluctuations, and one needs to take into account the fluctuation effects around the saddle point solutions. We thus make expansion of $\chi(\mathbf{q},i\nu_n)$ with respect to $i\nu_n$ and the momentum around the momentum loop. As shown in Fig.\ref{figadd1}(a), we define $\mathbf{p}=\mathbf{q}-\mathbf{Q}$, where $\mathbf{Q}$ lies on the momentum loop $S^1$ with radius $Q$. We are interested in the long-wave length fluctuations with $|\mathbf{p}|\rightarrow0$. Inserting  $\mathbf{q}=\mathbf{p}+\mathbf{Q}$ into Eq. \eqref{eqad20}, Eq.\eqref{eqad21} and making expansion with respect to $|\mathbf{p}|$ and $i\nu_n$, we obtain
\begin{equation}\label{eqad22}
  -S^{(2)}_{eff}=\sum_{\mathbf{q},i\nu_n}[-id\nu_n-a-b(|\mathbf{q}|-Q)^2]\Delta^{\dagger}(\mathbf{q},i\nu_n)\Delta(\mathbf{q},i\nu_n),
\end{equation}
where
\begin{equation}\label{eqad23}
a= \frac{2}{V}-2\sum_{\mathbf{k}}\frac{n_F(\epsilon_{v,\mathbf{k}}-\mu+D/2)-n_F(\epsilon_{c,\mathbf{k}+\mathbf{Q}}-\mu-D/2)}{\epsilon_{c,\mathbf{k}+\mathbf{\mathbf{Q}}}-\epsilon_{v,\mathbf{k}}-D}.
\end{equation}
and
\begin{equation}\label{eqad24}
\begin{split}
b&=\frac{1}{m}\sum_{\mathbf{k}}\frac{n_F(\epsilon_{v,\mathbf{k}}-\mu+D/2)-n_F(\epsilon_{c,\mathbf{k}+\mathbf{Q}}-\mu-D/2)}{(\epsilon_{c,\mathbf{k}+\mathbf{Q}}-\epsilon_{v,\mathbf{k}}-D)^2}\\
&-\frac{1}{m^2}\sum_{\mathbf{k}}\frac{n_F(\epsilon_{v,\mathbf{k}}-\mu+D/2)-n_F(\epsilon_{c,\mathbf{k}+\mathbf{Q}}-\mu-D/2)}{(\epsilon_{c,\mathbf{k}+\mathbf{Q}}-\epsilon_{v,\mathbf{k}}-D)^3}(k+Q)^2,
\end{split}
\end{equation}
and
\begin{equation}\label{eqad25}
d=2\sum_{\mathbf{k}}\frac{n_F(\epsilon_{v,\mathbf{k}}-\mu+D/2)-n_F(\epsilon_{c,\mathbf{k}+\mathbf{Q}}-\mu-D/2)}{(\epsilon_{c,\mathbf{k}+\mathbf{Q}}-\epsilon_{v,\mathbf{k}}-D)^2}.
\end{equation}
It is observed from Eq.\eqref{eqad23} that the requirement of $a=0$ is identical to the mean-field equation in Eq.\eqref{eqad16} at the critical point.

To the fourth order expansion of $\mathrm{ln}(-G^{-1})$ in  Eq.\eqref{eqad20}, we obtain
\begin{equation}\label{eqad26}
-S^{(4)}_{eff}=-c\sum_{q_1,q_2,q_3} \Delta^{\dagger}(\mathbf{q}_1,i\nu_{n,1})\Delta^{\dagger}(\mathbf{q}_2,i\nu_{n,2})\Delta(\mathbf{q}_3,i\nu_{n,3})\Delta(\mathbf{q}_1+\mathbf{q}_2-\mathbf{q}_3,i\nu_{n,1}+i\nu_{n,2}-i\nu_{n,3}),
\end{equation}
where the sum over $q$ includes both the momentums and frequencies, and
  \begin{equation}\label{eqad27}
c=2\sum_{\mathbf{k}}\frac{n_F(\epsilon_{v,\mathbf{k}}-\mu+D/2)-n_F(\epsilon_{c,\mathbf{k}+\mathbf{Q}}-\mu-D/2)}{(\epsilon_{c,\mathbf{k}+\mathbf{Q}}-\epsilon_{v,\mathbf{k}}-D)^3}.
\end{equation}
Since the higher-order terms of the bosons are more irrelevant, it is sufficient to  keep the effective action $S_{eff}=S^{(2)}_{eff}+S^{(4)}_{eff}$, and

\begin{equation}\label{eqad28}
\begin{split}
-S_{eff}&=\sum_{\tilde{\mathbf{q}},i\nu_n}[i\nu_n-\frac{a}{d}-\frac{b}{d}(|\tilde{\mathbf{q}}|-Q)^2]b^{\dagger}(\mathbf{q},i\nu_n)b(\mathbf{q},i\nu_n)\\
&-\frac{c}{d^2}\sum_{q_1,q_2,q_3} b^{\dagger}(\mathbf{q}_1,i\nu_{n,1})b^{\dagger}(\mathbf{q}_2,i\nu_{n,2})b(\mathbf{q}_3,i\nu_{n,3})b(\mathbf{q}_1+\mathbf{q}_2-\mathbf{q}_3,i\nu_{n,1}+i\nu_{n,2}-i\nu_{n,3}).
\end{split}
\end{equation}
From the action $S_{eff}$, we can read off the effective Hamiltonian describing the renormalized excitons:
\begin{equation}\label{eqad29}
H_{eff}=\sum_{\mathbf{q}}[\frac{(|\mathbf{q}|-Q)^2}{2m_b}-\mu_b]b^{\dagger}_{\mathbf{q}}b_{\mathbf{q}}+U\sum_{\mathbf{q}_1,\mathbf{q}_2,\mathbf{q}_3} b^{\dagger}_{\mathbf{q}_1}b^{\dagger}_{\mathbf{q}_2}b_{\mathbf{q}_3}b_{\mathbf{q}_1+\mathbf{q}_2-\mathbf{q}_3},
\end{equation}
where $\frac{1}{2m_b}=\frac{b}{d}$, $\mu_b=-\frac{a}{d}$, and $U=\frac{c}{d^2}$. It is known from the inset to Fig.2(e) of main text that $\chi(\mathbf{q})$ is an opening-down parabolic function of $|\mathbf{q}|$ around $Q$, indicating $b>0$ in Eq.\eqref{eqad22}, thus the effective mass $m_b>0$. With increasing the inter-band interaction between electrons $V$ (or lowering $\mu_0$), $a$ decreases following Eq.\eqref{eqad23}, and $\mu_b$ increases. For a critical $V_{cri}$ (or $\mu_{0,cri}$), $\mu_b$ crosses zero, suggesting a critical point where the excitons intend to condense. From Eq.\eqref{eqad23}, we also know that  this is the point where the critical mean-field self-consistent equation is satisfied.  However, Eq.\eqref{eqad29} clearly shows that at the critical regime, although excitons are formed, they feel a strong frustration as a moat band dispersion is spontaneously generated.

The above derivations reveal the relation between the boson theory in Eq.\eqref{eqad29} and the original fermion theory $H_0+H_I$. With varying the parameters $\mu_0$, $D$ or $V$ in the fermionic model, $\mu_b$, $m_b$ and $U$  of the bosons can be determined. Using the method introduced in  Sec. 3.1, we can obtain the trajectory of the boson parameters with varying the fermion parameters, as illustrated by Fig.2(d) of the main text. This leads to the full phase phase diagram shown in Fig.2(f) of the main text.

We mention that the transition between the electron (or hole) gas and the ETO is a topological phase transition beyond the Landau's framework of symmetry breaking. No other symmetries other than the time-reversal is broken when crossing the critical point. By analogy, the nature of the transition could be similar to the transition between the U(1) spin liquid with a spinon Fermi surface and the gapped Kalmeyer-Laughlin chiral spin liquid, which is an interesting research topic for future studies.
\subsection{3.4 Energetics of the excitonic topological order}
We now discuss several properties related to the energetics of the ETO state.

First, this state can be viewed as the composite fermion state that fully fills the lowest Landau level. The Landau level $E_l$ is obtained in Eq.\eqref{eq64}, as a  function of the intrinsic field $B_{CS}$. Since $B_{CS}$ is related to the density $n$ via $B_{CS}=2\pi n$, $E_l$ shows unusual dependence on $n$, as illustrated by Ref.\cite{Rui5s}. The larger the particle density $n$ is, the larger intrinsic field emerges, resulting in a larger degeneracy in each Landau level. As a result, the Landau level will be able to accommodate all the composite fermions. Therefore, the maximum number of the filled LLs can only be one in the ETO state, independent of $n$.

Second, we make a more careful comparison between the energy of the ETO state and the boson condensates. We use the parameters for the ETO state extracted from the experiments \cite{Rui5s}, namely, $\tilde{m}=1/(m^{-1}_e+m^{-1}_h)$, where $m_e\sim0.032m_0$, $m_h\sim0.136m_0$  and $m_0$ is the bare electron mass, $U\sim$1meV. Then, we let $n=xn_0$ where $n_0=Q^2/2\pi\sim1.43\times10^{10}$ cm$^{-2}$, with $x$ being the tuning parameter.  We show the calculated energies of the three different states with tuning $x$ in Fig.\ref{figadd2} below.
\begin{figure*}[t]
\includegraphics[width=0.6\linewidth]{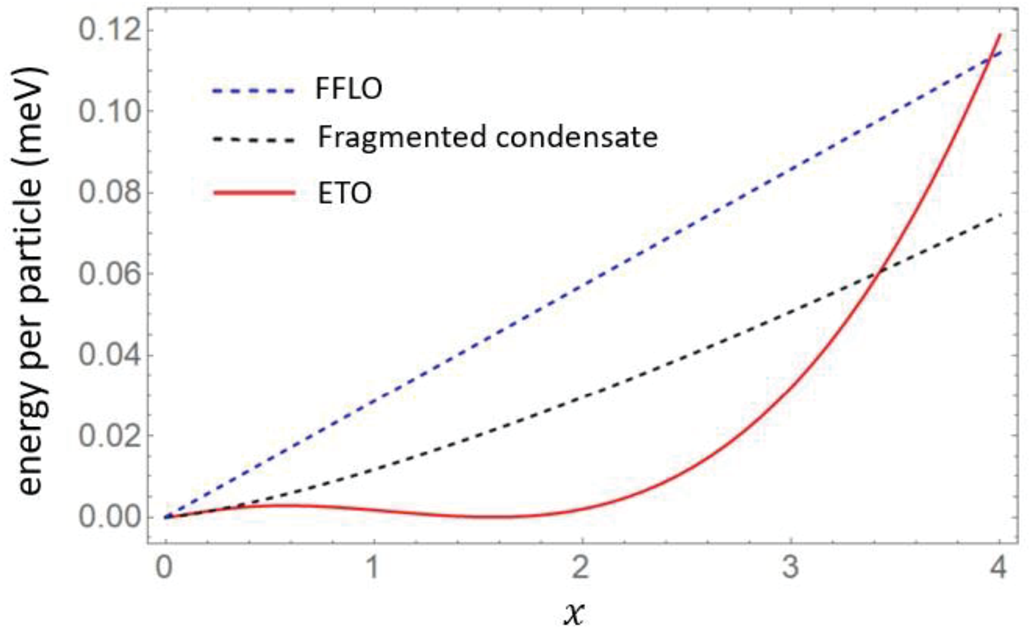}
\caption{Comparison of the energetics of the ETO, the fragmented boson condensate and the FF/LO state, with varying the particle density $n=xn_0$.}\label{figadd2}
\end{figure*}
As shown, for $x\lesssim4$, the FFLO state is ruled out in terms of energetics. The energies of the fragemented condensate (FC) and the ETO (chiral topological order) is close for $x\rightarrow0$ (the energy of ETO is still lower), however, the ETO has significantly lower energy in a wide region ($0.4\lesssim x\lesssim3.4$). At fixed density $n=2n_0$, the energy differences can be as large as, $E_{FC}-E_{TO}~0.03$meV and $E_{FFLO}-E_{TO}\sim0.06$meV. This clearly shows that there is a finite region where the chiral topological order becomes the ground state, as shown in Fig.2(f) of the main text.

Third, the excitation energy on top of the ETO state can be estimated by the spacing of LLs in terms of the composite fermions, i.e., $\Delta_{ETO}=E_1-E_0$, where $E_1$ and $E_0$ are the energy of the first excited state and the ground state respectively. Using Eq.\eqref{eq65} and the same parameters as above, $\Delta_{ETO}\sim 0.14$meV can be estimated. The ETO gap is much smaller than the gap of the excitonic insulator state, which is obtained around $3$meV using the same parameters in the mean-field calculations. This shows that the ETO state dominates the low-energy physics.

\subsection{3.5 Properties of the excitonic topological order}
The ETO is firstly revealed in the shallowly-inverted InAs/GaSb quantum well with a semimetal parent state \cite{Rui5s}. To maintain self-consistence, we briefly discuss in this subsection the properties of ETO.

The ETO wave function is given by Eq.\eqref{eq57}, Eq.\eqref{eqn106} and Eq.\eqref{eqn107}, i.e., the composite fermions filling the lowest Landau level. Moreover, Eq.\eqref{eq57} states that the composite fermions are essentially bosons (excitons in the ETO example) attached to 1-flux quanta. This state is clearly a chiral topological order. It can be found equivalent to $\nu=1/2$ bosonic (excitonic) fractional quantum Hall (FQH) state.

Let us start from an $\nu=1/2$ bosonic FQH under the magnetic field $B_{\perp}$, and $\nu=\phi_D\rho_0/B_{\perp}=1/2$, where $\phi_D$ is the flux quantum, and $\rho_0$ the number of particles. Then,  we regard a boson as an effective fermion attached to 1-flux quantum. In other words, 1-flux quantum $\phi_D$ is attached to a boson to yield a composite fermion. Then, the effective magnetic field $B_{eff}$ seen by the composite fermions is $B_{eff}=B_{\perp}-\Phi_D\rho_0=B_{\perp}/2$. Thus, the effective filling factor of composite fermions becomes $\nu_{eff}=\phi_D\rho_0/B_{eff}=1$. Therefore, the $\nu=1/2$ bosonic FQH can be viewed as $\nu_{eff}=1$ quantum Hall state of composite fermions, which are bosons attached to 1-flux quantum. Hence, the ETO state is in essence an $\nu=1/2$ excitonic FQH.

The effective field theory describing the ETO state is given by the Chern-Simons theory
\begin{equation}\label{eqada1}
  \mathcal{L}=\frac{k}{4\pi}\epsilon^{\mu\nu\rho}a_{\mu}\partial_{\nu}a_{\rho},
\end{equation}
with level $k=2$. This is because the Landau level contributes to a Chern-Simons term with coefficient 1 (since the Chern number is 1), and the statistical transmutation from fermions to bosons contributes to another Chern-Simons term with coefficient 1. Thus,  $k=2$  in total.

The above analysis and Eq.\eqref{eqada1}  indicate the following topological properties of the ETO state, namely, 1) the semionic excitations in the bulk, 2) the double degeneracy of ground state on a torus, and 3) the existence of chiral edge state. We mention that due to the excitonic nature, the chiral edge state has the inner structure consisting of electron and hole channels. Its experimental manifestation in the edge transport has been studied in Ref.\cite{Rui5s}. Clearly, this state is  similar in its topological nature to the Kalmeyer-Laughlin chiral spin liquid (CSL), as will be discussed in the next section.

\section{4. Chiral spin liquids in frustrated quantum magnets}
\subsection{4.1 2D Jordan-Wigner transformation}
Our theory can also predict chiral TO in frustrated quantum magnets, in particular the chiral spin liquids. In the main text, we propose a concrete quantum spin model, namely, the $J_1$-$J_2$-$J_3$ XY model on square lattice, which can satisfy the susceptibility condition. Here, we present the analytic and numerical details.

The $J_1$-$J_2$-$J_3$ XY model on square lattice reads as
\begin{equation}\label{eqaddn1}
  H=\sum_{\mathbf{r},\mathbf{r}^{\prime}}\frac{J_{\mathbf{r},\mathbf{r}^{\prime}}}{2}(S^x_{\mathbf{r}}S^x_{\mathbf{r}^{\prime}}+S^y_{\mathbf{r}}S^y_{\mathbf{r}^{\prime}}),
\end{equation}
where the sum over $\mathbf{r}$ and $\mathbf{r}^{\prime}$ evolves the first, second, and third nearest bond with the exchange interactions, $J_1$, $J_2$, and $J_3$. The model can be readily fermionized using the 2D Jordan-Wigner transformation \cite{Sedrakyan1s,Sedrakyan2s,Sedrakyan3s,Sedrakyan4s,Sedrakyan5s,Rui1s,Rui2s,Rui3s}. The spin raising/lowering operators for spin-1/2 systems can be equivalently mapped to spinless CS fermions coupled to string operators, i.e.,
\begin{eqnarray}\label{eqex0}
  S^{+}_{\mathbf{r}} &=& f^{\dagger}_{\mathbf{r}}e^{i\sum_{\mathbf{r}^{\prime}}\mathrm{arg}(\mathbf{r}-\mathbf{r}^{\prime})n_{\mathbf{r}^{\prime}}} \\
  S^{-}_{\mathbf{r}} &=& f^{\dagger}_{\mathbf{r}}e^{-i\sum_{\mathbf{r}^{\prime}}\mathrm{arg}(\mathbf{r}-\mathbf{r}^{\prime})n_{\mathbf{r}^{\prime}}}
\end{eqnarray}
where $f_{\mathbf{r}}$ ($f^{\dagger}_{\mathbf{r}}$) is the annihilation (creation) operator for the CS fermions, $\mathrm{arg}(\mathbf{r}-\mathbf{r}^{\prime})$ is the angle of the vector $\mathbf{r}-\mathbf{r}^{\prime}$,  and $n_{\mathbf{r}}=f^{\dagger}_{\mathbf{r}}f_{\mathbf{r}}$. Then, we obtain
\begin{equation}\label{eqaddn2}
  H=J_1\sum_{\mathbf{r},j}f^{\dagger}_{\mathbf{r}}f_{\mathbf{r}+\mathbf{e}_j}e^{iA_{\mathbf{r},\mathbf{r}+\mathbf{e}_j}}
  +J_2\sum_{\mathbf{r},j}f^{\dagger}_{\mathbf{r}}f_{\mathbf{r}+\mathbf{a}_j}e^{iA_{\mathbf{r},\mathbf{r}+\mathbf{a}_j}}
  ++J_3\sum_{\mathbf{r},j}f^{\dagger}_{\mathbf{r}}f_{\mathbf{r}+\mathbf{a}_j}e^{iA_{\mathbf{r},\mathbf{r}+\mathbf{b}_j}},
\end{equation}
where $A_{\mathbf{r},\mathbf{r}^{\prime}}=\sum_{\mathbf{r}^{\prime\prime}}\mathrm{arg}(\mathbf{r}-\mathbf{r}^{\prime\prime})n_{\mathbf{r}^{\prime\prime}}-\sum_{\mathbf{r}^{\prime\prime}}\mathrm{arg}(\mathbf{r}^{\prime}-\mathbf{r}^{\prime\prime})n_{\mathbf{r}^{\prime\prime}}$
is the lattice gauge field inherited from the fermionization, and $\mathbf{e}_j$, $\mathbf{a}_j$, $\mathbf{b}_j$ denote the lattice vectors of the first, second and third nearest neighbor bonds. The gauge fluctuations exhibit a Chern-Simons term, and can be exactly integrated out. This leads to a pure fermionic model, $H_0+H_I$, where
\begin{equation}\label{eqaddnew2}
  H_0=J_1\sum_{\mathbf{r},j}f^{\dagger}_{\mathbf{r}}f_{\mathbf{r}+\mathbf{e}_j}e^{i\overline{A}_{\mathbf{r},\mathbf{r}+\mathbf{e}_j}}
  +J_2\sum_{\mathbf{r},j}f^{\dagger}_{\mathbf{r}}f_{\mathbf{r}+\mathbf{a}_j}e^{i\overline{A}_{\mathbf{r},\mathbf{r}+\mathbf{a}_j}}
  ++J_3\sum_{\mathbf{r},j}f^{\dagger}_{\mathbf{r}}f_{\mathbf{r}+\mathbf{a}_j}e^{i\overline{A}_{\mathbf{r},\mathbf{r}+\mathbf{b}_j}},
\end{equation}
with $\overline{A}_{\mathbf{r},\mathbf{r}^{\prime}}$ the static gauge field that can be self-consistently determined by minimizing the energy of fermions. The gauge fluctuations further induce the fermion-fermion interaction,
\begin{equation}\label{eqaddnew3}
  H_I=\sum_{\mathbf{k}_1,\mathbf{k}_q,\mathbf{q}}v_{\mathbf{q}}f^{\dagger}_{\mathbf{k}_1}f_{\mathbf{k}_1+\mathbf{q}}f^{\dagger}_{\mathbf{k}_2}f_{\mathbf{k}_2-\mathbf{q}}.
\end{equation}
Eq.\eqref{eqaddnew2} and Eq.\eqref{eqaddnew3} are the fermion model obtained by exactly mapping from the original spin model.

We first consider the free sector $H_0$ with $\overline{A}_{\mathbf{r},\mathbf{r}^{\prime}}$ being  self-consistently determined. We focus on $J_3=J_2/2$ and gradually increase $J_2$ from 0 ($J_1$ is set to be the energy unit 1). For small $J_2$, the $\pi$-flux state is found to be stable, where Dirac fermions emerge in low-energy. The energy dispersion with increasing $J_2$ is shown by Fig.3(a)-(c) of the main text.

The bare susceptibility $\chi(\mathbf{q},0)$ only involves the Green's function of the free fermions. Hence,  $\chi(\mathbf{q},0)$ can be exactly calculated without any approximations. We calculate the susceptibility in the particle-hole channel, i.e., $\chi(\mathbf{r},\mathbf{r}^{\prime})=\langle\hat{T}\chi^{\dagger}_{\mathbf{r}}\chi_{\mathbf{r}^{\prime}}\rangle$ where $\chi_{\mathbf{r}}=f^{\dagger}_{\mathbf{r},-}f_{\mathbf{r},+}$, with ``$\pm$" here denoting the band index. This gives rise to
\begin{equation}\label{eqaddnew4}
  \chi(\mathbf{q},0)=-\sum_{\mathbf{k}}\frac{1-n_{\mathbf{k},-}-n_{\mathbf{k}+\mathbf{q},+}}{\epsilon_{\mathbf{k}+\mathbf{q},+}-\epsilon_{\mathbf{k},-}},
\end{equation}
where $n_{\mathbf{k}}$ is the Fermi distribution function. From Eq.(4), it can be found that the susceptibility displays the same maxima along a momentum loop, similar to that in Fig.2(d) of the main text. Thus, the susceptibility condition can be satisfied, suggesting emergence of chiral topological orders according to our theory.

Let us finally consider the intra-valley effects of the gauge-mediated interaction $H_I$. Clearly, the fermionized model shares the same low-energy physics with the ETO example discussed in the main text. For both small and large $J_2$, the fermion pairs condensate will be stable, leading to a long-range magnetically ordered state \cite{Sedrakyan5s,Rui2s}. While for intermediate $J_2$, the susceptibility condition is satisfied in the particle-hole channel, similar to that in the ETO example (Fig.2(a),(d) of the main text). Thus, our general theory predicts the occurrence of a chiral TO in an intermediate regime of $J_2$, which should be a chiral spin liquid in the context of the frustrated quantum spin systems.

\subsection{4.2 Numerical details}

Then, we use the tensor network renormalization group method to directly simulate the ground state of the frustrated $J_1$-$J_2$-$J_3$ XY model on a square lattice, which is defined by the following
\begin{equation}
	H = J_1\sum_{\langle i,j\rangle}\left(S^{+}_{i}S^{-}_j + h.c. \right) + J_2\sum_{\langle\langle i,j\rangle\rangle}\left(S^{+}_{i}S^{-}_j + h.c. \right) + J_3\sum_{\langle\langle\langle i,j\rangle\rangle\rangle}\left(S^{+}_{i}S^{-}_j + h.c. \right)
\end{equation}
where $J_1$, $J_2$, $J_3$ are the coupling constant between nearest, next-nearest, and next-next-nearest neighboring spins, respectively. In this work, we set $J_1 = 1$, $0.2\lesssim J_2 / J_1\lesssim 0.6$, and $ J_3 = J_2/2$.

\begin{figure*}[t]
	\includegraphics[width=0.4\linewidth]{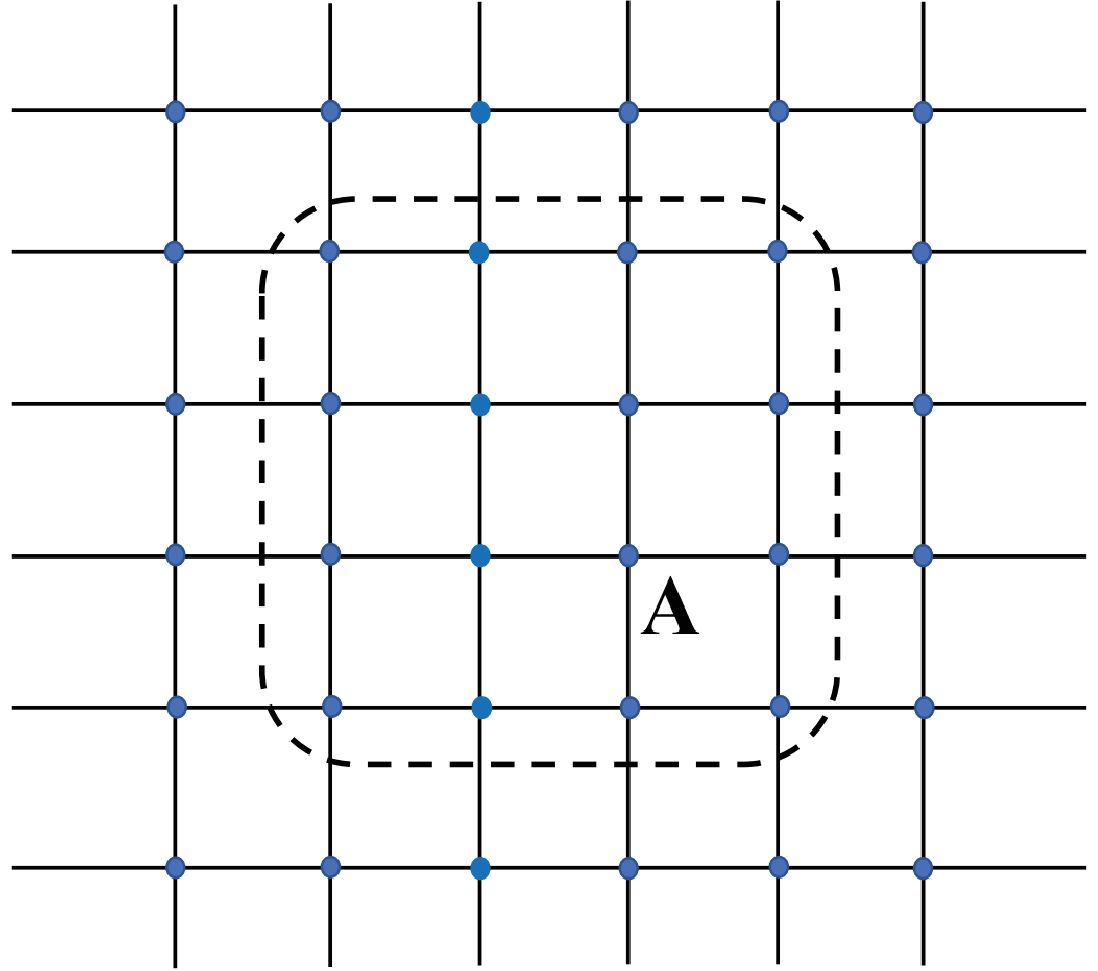}
	\caption{Tensor network state ansatz used in this work, corresponding to Eq.~(\ref{Wf}). The blue dots denote the local tensors $A^{\alpha,i}$s defined at the $i$-th site in the $\alpha$-th unit cell. The black dashed line denote the $4\times 4$ unit cell used in this work. Note that for simplicity, the physical index $\sigma_i$ is not shown in $A$ explicitly. In each time evolution step, the local tensor $A$ is used to characterize approximately the local entanglement among the five spins, i.e., the central spin and its four nearest neighbors, as explained in Ref.~\cite{PESS2014s} in detail.}\label{PESS}
\end{figure*}

From an arbitrary state represented as the projected entangled simplex state \cite{PESS2014s} defined in the following
\begin{equation}
	|\Psi_0\rangle = \sum_{\sigma}\mathrm{Tr}\left(\prod_{\alpha,i}A^{(\alpha,i)}_{l_ir_iu_id_i}[\sigma_i]\right)|\sigma\rangle \label{Wf}
\end{equation}
we employ imaginary-time evolution to obtained the approximated ground state.
In Eq.~(\ref{Wf}), the $\{\sigma\}$ denotes the spin configurations, $\mathrm{Tr}$ means summation over all the virtual indices $(lrud)$ of the local tensors $A^{\alpha,i}_{l_ir_iu_id_i}[\sigma_i]$s, each of which are defined at the $i$-th site in the $\alpha$-th unit cell. In this study, the unit cell is chosen as $4\times 4$, and the wave function is actually a 5-PESS and is very similar to the projected entangled pair state ansatz \cite{PEPSs}. To be specific, we follow the standard the simple update algorithm \cite{SU1D2007s, SU2D2008s} procedure for PESS wave function elaborated in Ref.~\cite{PESS2014s}. To further refine the state, we perform no more than 50 steps of full update \cite{FU2014s} after the representation is converged roughly. When the ground state is obtained, we employed the corner transfer-matrix renormalization group method \cite{CTMRG1996s, CTMRG2009s, CTMRG2014s} to contract the tensor network, in which the boundary dimension $\chi$ is kept no more than 120 so that the calculation can be performed effectively.
The PESS wave function is illustrated in Fig.~\ref{PESS}.

To study the chiral nature of the state, as done in Ref.~\cite{Rui3s}, we first wrap the obtained PESS wave function in an infinite long cylinder with circumference $L_y$ no more than 5 in the direction with periodic boundary condition, and then calculate the entanglement spectra corresponding to the bi-partition of the cylinder in the direction with open boundary condition. The idea to calculate the bi-partition entanglement spectra was elaborated in Ref.~\cite{Cirac2011s, Poil2016s, Saeed2018s} before. Regarding the two-dimensional network $\langle\Psi|\Psi\rangle$ as a trace of the power of some transfer matrices, we represent the dominant eigenvectors, $\sigma_L$ and $\sigma_R$ as matrix product states (MPS), determine them by boundary MPS method, and then represent $\sigma_L^{T}\sigma_R$ as a matrix product operator, from which the required entanglement spectra $\lambda$'s can be obtained. To identify the chiral edge mode with a bearable cost, in our calculation, we construct the Krylov-subspace (with dimension no more than 25) for a given lattice momentum $k$ (see, e.g., Ref. \cite{krylovs}), and determine the first $q$ largest eigenvalues in each $k$-space with $q\geq 10$.

It is known that the hyper-parameter, bond dimension $D$, controls the accuracy of the state representation and the total number of parameters one need to optimize by variation. Unfortunately, the memory and computational cost increases very fast with D, as discussed in detail in Ref.~\cite{NTNs}. Thus in practice one need to make a good balance between accuracy and cost. In this study, we have tried our best to push $D$ up to 10, which is believed be sufficient for our purpose.

%


\begin{thebibliography}{99}
\bibitem{footnote4} The fermion density equals to the boson density since the particle number operator remains unchanged in the transformation.
\end{thebibliography}
\end{document}